\documentclass[aps,twocolumn,superscriptaddress]{revtex4}
\usepackage{graphicx}
\usepackage{epsfig} % use this package to include EPS format figures
\usepackage{wrapfig}
\usepackage{amsmath}
\usepackage{amsfonts}
\usepackage{gensymb}

\usepackage{url}
\newcommand{\bfB}{{\mathbf{B}}}

\newcommand{\bfE}{{\mathbf{E}}}
\newcommand{\bfe}{{\mathbf{e}}}

\newcommand{\bfk}{{\mathbf{k}}}

\newcommand{\bfr}{{\mathbf{r}}}

\newcommand{\bfv}{{\mathbf{v}}}

\newcommand{\rmc}{{\mathrm c}}

\newcommand{\rmd}{{\mathrm d}}

\newcommand{\rme}{{\mathrm e}}

\newcommand{\rmDe}{{\mathrm{De}}}

\newcommand{\rmce}{{\mathrm{ce}}}

\newcommand{\rmpe}{{\mathrm{pe}}}
\newcommand{\rmpi}{{\mathrm{pi}}}

\newcommand{\cA}{{\mathcal{A}}}
\newcommand{\cE}{{\mathcal{E}}}
\newcommand{\cF}{{\mathcal{F}}}
\newcommand{\cK}{{\mathcal{K}}}
\newcommand{\cP}{{\mathcal{P}}}
\newcommand{\cR}{{\mathcal{R}}}
\newcommand{\cS}{{\mathcal{S}}}
\newcommand{\cT}{{\mathcal{T}}}
\newcommand{\cX}{{\mathcal{X}}}
\newcommand{\cY}{{\mathcal{Y}}}

\begin{document}

\def\Marseille{Aix-Marseille Universit{\'e}, CNRS, UMR 7345-PIIM, \\
                       case 322 campus Saint-J\'er\^ome, av.\ esc.\ Normandie-Niemen, 52,
                       FR-13397 Marseille cx 20, France}
\def\CEFIPRA{Indo-French Centre for the Promotion of Advanced Research-CEFIPRA, New Delhi, India}
\def\Nancy{Universit\'{e} de Lorraine, Institut Jean Lamour, UMR 7198, CNRS, France}
\def\CPT{Aix-Marseille Univ, Universit{\'e} de Toulon, CNRS, % UMR 7332 
               CPT, Marseille, France}
               %Aix Marseille Univ., Université de Toulon, CNRS, CPT, 13228 Marseille, France
\def\IPR{Institute for Plasma Research, Gandhinagar 382428, India}
               
\title{Sticky islands in stochastic webs and anomalous chaotic cross-field particle transport by E $ \times$ B electron drift instability}

\author{D. Mandal}
\email{debuipr@gmail.com}
\affiliation{\Marseille}
\affiliation{\CEFIPRA}
\affiliation{\IPR}

\author{Y. Elskens}
\email{yves.elskens@univ-amu.fr}
\affiliation{\Marseille}

\author{X. Leoncini}
\email{xavier.leoncini@univ-amu.fr}
\affiliation{\CPT}

\author{N. Lemoine}
\email{nicolas.lemoine@univ-lorraine.fr}
\affiliation{\Nancy}

\author{F. Doveil}
\email{fabrice.doveil@univ-amu.fr}
\affiliation{\Marseille}

%\begin{document}
%
\begin{abstract}
The $\bfE \times \bfB$ electron drift instability, present in many plasma 
devices, is an important agent in cross-field particle transport. 
In presence of a resulting low frequency electrostatic wave, 
the motion of a charged particle becomes chaotic and generates a stochastic 
web in phase space. We define a scaling exponent to characterise transport in 
phase space and we show that the transport is anomalous, of super-diffusive 
type. Given the values of the model parameters, 
the trajectories stick to different kinds of islands in phase space, 
and their different sticking time power-law statistics generate successive 
regimes of the super-diffusive transport. 

%This sticking time follows a power-law decay 
%$T_{\rm s}^{1-\gamma_{\rm Pk}}$, with exponent $\gamma_{\rm Pk}$ 
%for different sticking sets having different values in those successive 
%regimes. 
% and we observe 
%two different values of exponent  $\alpha = 0.17$ and $0.38$ 
%in the scaling *** (explain meaning of this $\alpha$).    
%For a particular set of parameter we found a Halloween mask like web structure.

\bigskip


   52.20.Dq   Particle orbits \\
   52.25.Fi    Transport properties \\
   52.75.Di    Ion and plasma propulsion  \\
   05.45.Pq   Numerical simulations of chaotic systems \\

\end{abstract}

%\bigskip \date{\today{} \copyright The authors  --1st Rev, Version 011dm -- DRAFT in progress}

\maketitle

%%%%%%%%%%%%%%%%%%%%%%%%%%%%%%%%%%%%%%%%%
\section{Introduction}
\label{sec:Intro}
%%%%%%%%%%%%%%%%%%%%%%%%%%%%%%%%%%%%%%%%%

The formation of stochastic web structures and the chaotic transport of charged 
particles in presence of electrostatic waves and magnetic field have been 
investigated for several decades \cite{zaslavsky91, chernikov87, vasilev91, 
balescu05, bouchara15, chen01, benkadda96}. 
In purely chaotic situations where a central limit theorem is valid, 
the transport process is like a discrete time random walk, and the variance 
grows linearly with time \cite{vkampen92, balescu97, oeksendal03}. 
But in the case of mixed phase space where both chaotic and regular 
trajectories coexist, the transport processes are not so clear 
\cite{Chirikov91, Bunimovich15, Lozej18}. 
Usually, trajectories spend more time near the
border of the regular region. This type of dynamics can sometimes be modeled 
using continuous time random walks (CTRW), where the number of jumps
within a time interval $[0,t)$, and the displacement in each jump 
are taken from two mutually independent probability densities, 
and these two probability densities fully specify the probability distribution 
describing the random walk \cite{Balescu:r,Mendez:v}. Transport in such systems 
can be linked with L\'evy flight type processes \cite{shlesinger95, Negrete:d, Solomon:t}.
In presence of a magnetic field, due to the interaction with electrostatic 
waves, the dynamics of the charged particles become chaotic and, for certain 
parameter values, they form stochastic webs where chaotic sticky islands, 
inside which trajectories show regular features, coexist with a chaotic 
``sea'' between islands. Large scale transport is possible through this 
chaotic domain \cite{Boozer94, zaslavsky07, contopoulos10}. 

These web structures exhibit different shapes which depend 
on the wave vectors $\bfk$ and amplitudes of the electrostatic waves,  
and on $\omega_\bfk / \omega_\rmce$ the frequency ratios
of electrostatic waves frequencies $\omega_\bfk$ 
to the electron cyclotron frequency $\omega_\rmce$ \cite{karney78}. 
The study of particle transport in these web structures helps 
to understand the anomalous collisionless transport mechanism in magnetized plasmas. 
In most previous studies, the formation of stochastic webs and 
the associated transport were investigated for high wave frequency 
($\omega_\bfk \gg \omega_\rmce$). 
Moreover, the perpendicular diffusion coefficient in that limit is calculated 
by invoking the linear time dependence of the variance \cite{karney79}. 
However, deviations from the linear time dependence of variance are
frequently observed in case of mixed phase space.

Here, we consider the collisionless transport mechanism of electrons 
due to the $\bfE \times \bfB$ electron drift instability. 
In magnetized Hall plasmas \cite{kaganovich20}, the $\bfE \times \bfB$ electron drift, 
plasma density, temperature, magnetic field gradients and ion flow 
are the sources of the $\bfE \times \bfB$ drift instabilities or
electron cyclotron drift instability \cite{mikhailovskii92}. 
This instability is observed in many magnetized plasma devices 
like magnetrons for material processing \cite{abolmasov12}, 
magnetic filters \cite{boeuf12}, Penning gauges \cite{ellison12}, 
linear magnetized plasma devices dedicated to study cross-field plasma 
instabilities \cite{matsukuma03}, Hall thrusters for space propulsion 
and many fusion devices. In Hall thrusters and other devices, 
this $\bfE \times \bfB$ drift instability plays a dominant role in 
anomalous particle transport. 
In most of these devices, the electrostatic modes generated by 
$\bfE \times \bfB$ drift instability have very small frequencies compared to 
the electron cyclotron frequency ($\omega_\bfk  \ll  \omega_\rmce$). 
Therefore, the resonance condition with the cyclotron harmonics, 
$\omega_\bfk - k_\parallel v_\parallel = \ell \omega_\rmce$, is not satisfied.  

In previous works, 
the chaotic dynamics of test particle (electron) near the anode region in Hall 
thrusters due to inhomogeneities in magnetic field, and its effect 
on ionization efficiency and anomalous electron transport, were reported 
even in the absence of waves \cite{Marini:S,Skovoroda:A}. 
In our recent work \cite{mandal19a}, we present the anomalous transport 
of electrons due to wave-particle interaction in Hall thruster 
using a three-dimensional test particle model, 
even in a uniform magnetic field. 
But in three space dimensions,
the dynamics in the presence of waves is very complicated. 

In this paper, we focus on the consequence of the $\bfE \times \bfB$ 
drift instability. 
We discuss the formation of chaotic web structures 
and characterize the associated transport properties 
using a reduced two-degrees-of-freedom Hamiltonian 
which helps to simplify the original dynamics complexity. 
In real thrusters, due 
to the presence of wall boundaries, particles can reflect from the boundary, 
a process which may destroy the web formation. We find that, 
in presence of a single background electrostatic wave 
along with the uniform, static electric and magnetic fields, 
the trajectories generate web structures, 
and that, due to the presence of sticky islands, particle transport is super-diffusive. 
Then we analyse the sticking time statistics by observing 
a power-law decay of particle presence in each of the sticky sets. 
Therefore, this work complements our previous findings 
with a new understanding of the particle transport, 
which can be applicable in other systems having mixed 
phase space.

Section~\ref{sec:Model} presents the model and its two descriptions 
(respectively time-dependent and time-independent). 
Sec.~\ref{sec:Numeric} indicates the numerical method used to integrate the evolution equations. 
Sec.~\ref{sec:web} discusses the chaotic web structures 
generated by the dynamics, 
Sec.~\ref{sec:Transport} analyses the transport in these structures,
and Sec.~\ref{sec:sticky_island} discusses the effect of sticking to invariant islands on transport. 
We conclude in Sec.~\ref{sec:Conclusion}.
%

%%%%%%%%%%%%%%%%%%%%%%%%%%%%%%%%%%%%%%%%%%%%
\section{Reduced Hamiltonian dynamics and the elementary model}
\label{sec:Model}
%%%%%%%%%%%%%%%%%%%%%%%%%%%%%%%%%%%%%%%%%%
%%%%%%%%%%%%%%%%%%%%%%%%%%%%%%%%%%%%%%%%%
\subsection{Fields acting on an electron}
\label{sec:Fields}
%%%%%%%%%%%%%%%%%%%%%%%%%%%%%%%%%%%%%%%%%
%
We consider a Cartesian coordinate system for the numerical modelling, 
with $x$-direction along the magnetic field $\bfB_0$,
$y$-direction as $\bfE_0 \times \bfB_0$ drift direction 
and $z$-direction along the constant electric field $\bfE_0$ 
(see fig.~ 1 of \cite{mandal19a}).

In Hall thruster geometry, unstable low frequency 
($\omega_{\bfk} \ll \omega_\rmce$) electrostatic waves 
are generated due to $\bfE \times \bfB$ drift instability. 
A 3D dispersion relation of this instability for Hall thruster has been derived by
Cavalier \emph{et al.} \cite{cavalier13}. 
The most unstable mode \cite{boeuf18} is given by
$k_{\max}  \sim (\lambda_{\rmDe} \sqrt{2})^{-1}$ and 
$\omega_{\max}  \sim \omega_{\rmpi} / \sqrt{3}$, where $\lambda_{\rm De}$ and
$\omega_{\rm pi}$ are the electron Debye length and ion plasma frequency respectively. 
Its propagation angle deviates by 
${\mathrm{tan}}^{-1}(k_z/k_y) \sim 10 - 15^\circ$ 
from the $y$-direction near the thruster exit plane. 
Hence, the wave vector along the $z$-direction is small ($k_z \sim 0.2 \, k_y$), 
and the electric field along the $z$-direction is dominated 
by the stronger constant axial electric field $\bfE_0$. 
Therefore, for simplicity, we remove here the $z$-variation of the electric 
field. 
%
%Then, the time varying part of the potential in $x-y$ plane is 
%constructed as a sum of unstable modes. 

For this first investigation, %for the numerical study, 
we consider only the fastest growing mode. 
The total electric field acting on the particle is
\begin{equation} 
  \bfE(x,y,z,t) 
  =  
  \phi_1 \bfk \sin \alpha(x,y,t)  
%  \nonumber \\ 
%  && 
  +  E_0 \, \bfe_z,
\label{e-field}
\end{equation} 
with the local phase $\alpha(x, y, t) :=  k_x  x +  k_y  y - \omega_1 t$, 
where the wave vector $\bfk = k_x \bfe_x + k_y \bfe_y$ 
and the wave angular frequency $\omega_1$ follow the dispersion relation of the 
$\bfE \times \bfB$ electron drift instability \cite{cavalier13} and $k_z = 0$.
The origin of time is such that $\alpha = 0$ for $x=y=0$, $t=0$. 
The position $\bfr = (x, y, z)$, velocity $\bfv$, 
time $t$ and the potential $\phi_1$ are normalized with Debye length 
$\lambda_{\rmDe}$, thermal velocity $v_{\rm the}$, 
inverse electron plasma frequency 
$\omega_{\rm pe}^{-1}$ and $m_\rme v_{\rm the}^2 / | q_\rme |$, respectively.
We choose the amplitude $\phi_1$ equal to the saturation potential 
\cite{boeuf18} at the exit plane of the thruster
$\vert \delta \phi_{y,{\rm rms}}\vert = T_\rme / (6 \sqrt{2}) 
= 0.056 \, v_{\rm the}^2$. 
We consider a single mode  
with ($k_{x}, k_{y}, \omega_1) = (0.001, 0.754, 1.23 \cdot 10^{-3})$. 

Earlier studies of Hall thruster \cite{cavalier13} reported that the comb-like unstable modes 
found in axial-azimuthal plane get smoothed out 
as one increases the wave vector $k_x$ parallel to the magnetic field~; 
for large $k_x$ values ($k_x > 0.08$),  
the dispersion relation tends to follow an asymptotic curve corresponding to 
a modified ion acoustic instability. 
At small $k_x$ values, the real part of the frequency given by the dispersion function oscillates
about this asymptotic curve and crosses this asymptotic curve at each resonance 
(and also one time between each resonance), as can be seen in figure 2 of ref.\ \cite{cavalier13}, 
while the growth rate depends strongly on $k_x$ at that point. 
As our model is not self-consistent, 
so that the amplitude of the wave is fixed according to experimental 
measurements in Hall thruster\cite{tsikata10}, 
we only use the real part of the frequency of the unstable mode. 
The mode considered is the one having the largest growth rate 
and corresponds to one of the resonances. 
So the real part of the frequency at that point is correctly given by the 
approximated dispersion relation \cite{boeuf18}, 
even though the value of $k_x$ chosen here is too small to ensure a smoothed 
dispersion relation.

%At small $k_x$ values, the dispersion 
%function oscillates about this asymptotic curve. 
%Indeed, in some 1D and 2D particle in cell (PIC) simulations 
%of $E\times B$ electron drift instability in condition of Hall thruster \cite{lafleur:t,boeuf18,Adam:j}, 
%the smoothing out of the electron cyclotron modes and the transition to an ion acoustic instability are observed, 
%even though the parallel component of the wave vector is zero in those models. 
%Therefore, at small value of $k_x$ also one can consider that, on average, 
%the dispersion relation follows approximately the modified ion-acoustic instability. 
%This motivates us to consider the simplified dispersion relation 
%of modified ion-acoustic instability for very small value of $k_x$.
%
From here on, we write $\omega_\rmc$ for $\omega_\rmce$. 
In normalized units, the cyclotron frequency 
$\omega_\rmc = |q_\rme B_0| / m_\rme = 0.1 \, \omega_{\rm pe}$, 
$|q_\rme E_0| / m_\rme = 0.04 \, \omega_{\rm pe} v_{\rm the}$, 
and the drift velocity $v_\rmd = E_0 / B_0 = 0.4 \, v_{\rm the}$.
Therefore, the $y$-component of the mode phase velocity is small
($\omega_1 / k_y \ll v_\rmd$).

% *** Is the fastest growing mode such that $k_x/k_y \sim 0.0013$ ? Why is it so small ? 

As a result, in the Lorentz equation of motion of a particle with mass $m$ and 
charge $q$
\begin{eqnarray}
  \ddot{\bfr} = \frac{q}{m} \left(\bfE(\bfr, t) + \dot{\bfr} \times \bfB \right),
\label{H_eqm}
\end{eqnarray}
the electric field $\bfE(\bfr, t)$ has a constant part $\bfE_0$ along 
$z$-direction and a slowly time varying part in $(x, y)$ plane. 
Eq.~(\ref{H_eqm}) can be written componentwise, 
using Eq.~(\ref{e-field}), as
\begin{eqnarray}
  \ddot{x} 
  &=& \frac{q E_{1x}}{m} \sin(k_x x + k_y y - \omega_1 t), 
  \label{x_eq}
  \\
  \ddot{y} 
  &=& \frac{q E_{1y}}{m} \sin(k_x x + k_y y - \omega_1 t) 
          + \omega_\rmc \dot{z}, 
  \label{y_eq}
  \\
  \ddot{z} 
  &=& \frac{q E_0}{m} - \omega_\rmc \dot{y}
 \label{z_eq} ,
\end{eqnarray}
where $E_{1x} = k_x \phi_1$ and $E_{1y}= k_y  \phi_1$ are the amplitude 
of the $x$- and $y$-components of electric field, respectively. 
%while $\omega_\rmc = q B_0 / m$ and $\omega_1$ 
%are the cyclotron and wave frequency, respectively. 
Eq.~(\ref{z_eq}) can be integrated:
\begin{equation}
  \dot{z} + \omega_\rmc  y = \frac{q E_0}{m} t + a, 
\end{equation}
where $a = v_{z0} + \omega_\rmc y_0$ is a constant of integration, 
$v_{z0}$ and $y_0$ are the particle's initial $z$-component velocity and 
position along $y$-direction, respectively. 
Substituting $\dot{z}$ in Eq.~(\ref{y_eq}), 
and recalling the drift velocity $v_\rmd = E_0 / B_0$,
we reduce the equation of motion of the particle 
to a system of two equations,
\begin{eqnarray}
\begin{aligned}
  \ddot{y} + \omega_\rmc ^2 y 
  & =  \frac{q E_{1y}}{m} \sin\left(k_x x + k_y y - \omega_1 t\right) 
           + v_\rmd \omega_\rmc ^2 t + \omega_\rmc a ,
  \\
  \ddot{x} 
  & =  \frac{q E_{1x}}{m} \sin\left(k_x x + k_y y - \omega_1 t\right).
\label{2D_Eqm}
\end{aligned}
\end{eqnarray}

%*** check everywhere : is there a + or a -- sign in front of the sine in the equations for $\ddot x, \ddot y$ ? 
%This will also fix the sign in front of the cosine in the hamiltonian ***

%%%%  hamiltonian viewpoint
%%%%
%%%%  we start from the lagrangian 
%%%%  \begin{eqnarray}
%%%%    L 
%%%%    & = & \frac{m}{2} (v_y^2 + v_z^2) + q v_z B_0 y - q E_0 z 
%%%%    \\
%%%%    & = & \frac{m}{2} (v_y^2 + v_z^2 + \omega_\rmc v_z y - \omega_\rmc d_\rmd z)
%%%%  \end{eqnarray}
%%%%  
%%%%  Conjugate momenta
%%%%  \begin{eqnarray}
%%%%    p_y & = & m v_y
%%%%    \\
%%%%    p_z & = & m v_z + m \omega_\rmc y
%%%%  \end{eqnarray}
%%%%  
%%%%  hamiltonian
%%%%  \begin{equation}
%%%%    H = \frac{p_y^2 + p_z^2}{2 m} + \frac{m}{2} \omega_\rmc^2 y^2 
%%%%           - \omega_\rmc y p_z + m \omega_\rmc v_\rmd z
%%%%  \end{equation}

%%%%%%%%%%%%%%%%%%%%%%%%%%%%%%%%%%%%%%%%%
\subsection{Time-dependent Hamiltonian}
\label{sec:Ham-time-dep}
%%%%%%%%%%%%%%%%%%%%%%%%%%%%%%%%%%%%%%%%%
%
System (\ref{2D_Eqm}) derives from the Hamiltonian $H(p_x, p_y, x, y, t)$
\begin{eqnarray}
  H 
  & = & 
  \frac{p_x^2 + p_y^2}{2m} 
  + \frac{m}{2} \omega_\rmc^2 \, y^2 - \left(t + A\right) m v_\rmd \omega_\rmc^2\, y
  \nonumber \\ 
  && + q \phi_1 \cos\left(k_x x + k_y y - \omega_1 t\right),
  \end{eqnarray}
where 
$A = \left(v_{z0} + \omega_\rmc y_0\right) / \left(\omega_\rmc v_\rmd\right)$ 
is a constant.  
By means of the generating function
\begin{equation} 
  F\left(P_x, P_y, x, y, t\right) 
  = P_x x + \left(P_y + v_\rmd\right) \left(y - (t+A) v_\rmd\right),
\label{Gen_F}
\end{equation} 
we change to new variables $(P_x, P_y, X, Y)$ in a frame  
moving with a constant velocity $v_\rmd$ along the $y$-direction
(which we call a ``{\it drifted frame}'' for figures),  
\begin{eqnarray}
\begin{aligned}
  X = \frac{\partial F}{\partial P_x}
  & = x, 
  \\
  Y = \frac{\partial F}{\partial P_y} 
  & = y - \left(t+A\right) v_\rmd, 
  \\
  p_x = \frac{\partial F}{\partial x} 
  & = P_x,
  \\
  p_y = \frac{\partial F}{\partial y} 
  & = P_y + v_\rmd, 
  \\
  \frac{\partial F}{\partial t} 
  & = - \left(P_y + v_\rmd\right) v_\rmd .
\label{trans_co}
\end{aligned}
\end{eqnarray}
Using these new coordinates (\ref{trans_co}), 
the new Hamiltonian (after removing terms irrelevant to the motion) 
and the equations of motion read 
\begin{eqnarray}
\begin{aligned}
  K\left(P_x, P_y, X, Y, t\right) 
   = \, & \frac{P_x^2 + P_y^2}{2m} 
         + \frac{m}{2} \omega_\rmc ^2 \, Y^2 
      %\\  & 
      + q \phi_1  \cos \alpha , 
   \\
   \ddot{X} 
   = \, & \frac{q \phi_1}{m} k_x \sin \alpha , 
   \\
   \ddot{Y} + {\rm \omega_\rmc ^2}Y 
   = \, & \frac{q \phi_1}{m} k_y \sin \alpha ,
\label{New_Hamil}
\end{aligned}
\end{eqnarray}
with $\alpha = k_x X + k_y Y + (v_\rmd k_y - \omega_1) t + \zeta$ and 
where $\zeta = k_x v_\rmd A$ is constant. 

The dimensionless equations of motion are obtained using the 
dimensionless variables $X' = k_x X + \zeta$, $Y' = k_y Y$,
$t' = \omega_\rmc t$. 
Introducing the new notation  $\beta = k_x / k_y$, 
$\varepsilon = q \phi_1 k_y^2 / (m \omega_\rmc ^2)$ and
$\nu_1 = (v_\rmd k_y - \omega_1) / \omega_\rmc $, 
we obtain the dimensionless equations 
\begin{eqnarray}
\begin{aligned}
  \frac{\rmd^2 X'}{\rmd t'^2}
  &= \varepsilon \beta^2 \sin \left( X' + Y' + \nu_1 t' \right), 
  \\
  \frac{\rmd^2 Y'}{\rmd t'^2} + Y' 
  &= \varepsilon \sin \left( X' + Y' + \nu_1 t' \right).
\label{TimeDep_Red}
\end{aligned}
\end{eqnarray}
In this paper, we solve Eqs~(\ref{TimeDep_Red}) numerically 
using a second order symplectic scheme. 
The dynamics involves two degrees of freedom with a time-periodic dynamics 
(with period $2 \pi / \nu_1)$, 
so that the effective phase space is 5-dimensional. 
The coordinate $X'$ admits periodic boundary condition (with period $2 \pi$), 
whereas $Y'$ runs over the real line. 

The dynamics depends on three parameters, $\varepsilon$, $\beta$ and $\nu_1$. 
For $\varepsilon = 0$, $X'$ is ballistic and $Y'$ is a harmonic oscillator, 
in agreement with the well-known solutions for particle motion 
in stationary, uniform fields $\bfE_0$ and $\bfB_0$. 

Note that, in Hall thrusters, $\bfB_0$ is radial and $\bfE_0$ is axial, 
so that the drift is azimuthal. The coordinates $y$ and $Y$ are thus defined on circles,
while $x$ and $X$ are actually bounded by the inner and outer cylindrical chamber walls. 
The origin for $Y$ and $X$ are determined by the initial conditions $(y_0, v_{z0})$ 
and the phase convention for the electrostatic mode, respectively. 

%%%%%%%%%%%%%%%%%%%%%%%%%%%%%%%%%%%%%%%%%
\subsection{Time-independent Hamiltonian}
\label{sec:Ham-time-indep}
%%%%%%%%%%%%%%%%%%%%%%%%%%%%%%%%%%%%%%%%%
%
A time-independent Hamiltonian can be derived by 
means of a Galileo transformation along $X$ with velocity $\nu_1 \omega_\rmc / k_x$. 
With the generating function and change of variables
\begin{eqnarray}
\begin{aligned}
  \cF 
  & = \left(\cP_x - \frac{\nu_1 m \omega_\rmc}{k_x}\right) 
     \left(X + \frac{\nu_1 \omega_\rmc t + \zeta}{k_x}\right) + \cP_y Y,
  \\
  \cX & = X + \frac{\nu_1 \omega_\rmc}{k_x} t + \frac{\zeta}{k_x}, 
  \\
  P_x & = \cP_x - \frac{\nu_1 m \omega_\rmc }{k_x}, 
  \\
  \cY & = Y, 
  \\
  P_y & = \cP_y, 
  \\
  \frac{\partial \cF}{\partial t} & = 
  \frac{\nu_1 \omega_\rmc }{k_x} \left(\cP_x - \frac{\nu_1 m \omega_\rmc}{k_x}\right) ,
\end{aligned}
\end{eqnarray}
the Hamiltonian (up to terms irrelevant to the motion) and the equations of motion can be written as 
\begin{eqnarray}
\begin{aligned}
  \cK \left(\cP_x, \cP_y, \cX, \cY\right) 
  & = & \frac{\cP_x^2 + \cP_y^2}{2 m} + 
    \frac{m}{2} \omega_\rmc ^2 \cY^2 
    \qquad \\ & & 
    + q \phi_1 \cos\left( k_x \cX + k_y \cY \right), 
  \\
  \ddot{\cX} 
  & = & \frac{q \phi_1}{m} k_x \sin\left( k_x \cX + k_y \cY \right), 
  \\
  \ddot{\cY} + {\rm \omega_\rmc ^2}\cY 
  & = & \frac{q \phi_1}{m} k_y \sin\left( k_x \cX + k_y \cY \right).
\label{New_Hamil2}
\end{aligned}
\end{eqnarray}
Setting $\cX' = k_x \cX$, $\cY' = k_y \cY$ and $t' = \omega_\rmc t$, 
the equations of motion (\ref{New_Hamil2}) reduce to
\begin{eqnarray}
  \begin{aligned}
    \frac{\rmd^2 \cX'}{\rmd t'^2} 
    &= \varepsilon \beta^2 \sin\left( \cX' + \cY' \right), 
    \\
    \frac{\rmd^2 \cY'}{\rmd t'^2} + \cY' 
    &= \varepsilon \sin\left( \cX' + \cY' \right).
  \label{Auton_Red}
  \end{aligned}
\end{eqnarray}
Eq.~(\ref{Auton_Red}) is solved numerically for various parameters and initial conditions.

This dynamics depends on two parameters only, $\varepsilon$ and $\beta$. 
It involves two degrees of freedom, 
with the coordinate $\cX'$ obeying periodic boundary condition 
(with period $2 \pi$), whereas $\cY'$ runs over the real line. 
As the dynamics is autonomous, it preserves the ``energy'' $\cK$. 
Therefore, trajectories stay on smooth 3-dimensional surfaces, 
and they may be visualised by means of 2-dimensional Poincar\'e sections. 

While coordinates $(x, y)$ and $(X, Y)$ are related with the Hall thruster chamber, 
coordinates $(\cX, \cY)$ simplify further the dynamics, 
provided one does not worry about boundary conditions. 
Therefore, we use both the time-independent and the time-dependent representations
in the following discussions. 

For $\varepsilon = 0$, viz.\ in absence of electrostatic wave, 
the dynamics is integrable. 
Its dimensionless actions are the linear momentum $\rmd \cX' / \rmd t'$ 
with angle the position $\cX'$ periodic with period $2 \pi$ in agreement with the boundary condition,
and the gyration energy ${\cR'}^2 / 2 = ({\cY'}^2 + (\rmd \cY' / \rmd t')^2) / 2$ 
(divided by the cyclotron frequency, which is 1)
with angle the gyrophase in the $(\cY', \rmd \cY' / \rmd t')$ plane. 
In presence of the electrostatic wave, for small $\varepsilon$,  
these actions generate two adiabatic invariants. 
For $\beta$ also small, the actions evolve on different time scales 
(in terms of the dimensionless $t'$),
namely $\varepsilon^{-1}$ for the oscillations of $\cY'$
and $\varepsilon^{-1} \beta^{-2}$ for the nearly-ballistic motion of $\cX'$.

%%%%%%%%%%%%%%%%%%%%%%%%%%%%%%%%%%%%%%%%%%%%%%
\section{Numerical method}
\label{sec:Numeric}
%%%%%%%%%%%%%%%%%%%%%%%%%%%%%%%%%%%%%%%%%%%%%%

%From here on, we omit all primes in the evolution equations (\ref{TimeDep_Red}) and (\ref{Auton_Red}).

Because the right hand sides of Eqs~(\ref{TimeDep_Red}) and (\ref{Auton_Red}) 
depend on space, the infinitesimal generators for both velocity and position 
equations do not commute, and one uses a time-splitting numerical integration 
scheme. Since the dynamics is Hamiltonian and we are interested in long-time evolution,
we choose a symplectic scheme, 
which guarantees preservation of the hamiltonian structure exactly 
and ensures over very long time the conservation of a hamiltonian close to the original one \cite{Benettin99,Hairer03}. 
%From here on, we omit all primes in the evolution equations 
%(\ref{TimeDep_Red}) and (\ref{Auton_Red}).

The positions are advanced with the map
$\bfr (t + \Delta t) = \cT_{v, \Delta t} (\bfr(t)) =  
\bfr (t) + \bfv \Delta t$, and
the velocities are advanced in the form
$ v_x (t + \Delta t) = \cT_{\cE x, \Delta t} (v_x(t)) 
= v_x (t) + \varepsilon \beta^2 \sin(\cX + \cY ) \, \Delta t$ and
$v_y (t + \Delta t) = \cT_{\cE y, \Delta t} (v_y(t)) 
= v_y (t) + (\varepsilon \sin(\cX +\cY ) - \cY) \, \Delta t$.
As a result, we use a second-order symmetric leapfrog scheme, 
which evolves (\ref{Auton_Red}) as the map
\begin{equation}
  \left( \begin{array}{c} \bfr(t + \Delta t) \\ 
          \bfv(t + \Delta t) \end{array} \right) 
  = \cA
  \left( \begin{array}{c} \bfr(t) \\ 
          \bfv(t) \end{array} \right) ,
  \label{matrix_oper}
\end{equation}
\begin{equation}
  \cA = \cT_{v, \Delta t/2} \circ  \cT_{\cE, \Delta t} \circ \cT_{v, \Delta t/2}  .
\label{Sympl_opert}
\end{equation}

We evolve Eqs~(\ref{TimeDep_Red}) similarly,
with $\cT_{\cE, \Delta t}$ evaluated at midstep $t + \Delta t/2$.

%%%%%%%%%%%%%%%%%%%%%%%%%%%%%%%%%%%%%%%%%
\section{Stochastic web structure}
\label{sec:web}
%%%%%%%%%%%%%%%%%%%%%%%%%%%%%%%%%%%%%%%%%

The values of $(\varepsilon, \beta^2, \nu_1)$ in a Hall thruster device 
for the fastest growing mode ($k_{x} = 0.001, k_y = 0.754, 
\omega = 1.23 \cdot 10^{-3} $)
are $\varepsilon =3.21$, $\beta^2 = 1.75 \cdot 10^{-6}$ and $\nu_1 = 3$. 
As $\varepsilon$ is not small, the gyration action is definitely not conserved, 
as will be seen in the phase space plots. 
However, $\varepsilon \beta^2 \sim 5 \cdot 10^{-6}$ 
so that the ballistic action is almost conserved over times 
$t \sim 10^5 \, \omega_{\rmc}^{-1} = 10^6 \, \omega_{\rmpe}^{-1}$.
Moreover, $k_x = 0.01$, so that $\varepsilon \beta^2 = 5.6\cdot 10^{-4}$ 
which also ensures conservation of  the action for long time. 
We checked that setting $k_x = 0.01$ with all other parameter kept constant 
gives similar phase space plots as  $k_x = 0.001$. 
For higher values like $k_x \ge 0.1$, the action is not conserved over long times, 
which changes the phase space structure.

Here we first focus on transport for web structures 
with three-fold rotational symmetry ($\nu_1 = 3$) as in the Hall thruster 
geometry, and its harmonic the six-fold rotational symmetry ($\nu_1 = 6$)
with respect to gyration variables $(Y', v_y)$. 
By ``\textit{rotational symmetry}'' we mean that, 
if one rotates the $(Y', v_y)$-space web structure by angle $2 \pi / n$ around the origin, then 
it generates an identical phase space structure.
To assess the importance of having an integer value for the forcing frequency 
$\nu_1$, we also consider the non-resonant value $\nu_1 = 1.39$. 
For the time-independent description, the initial velocity $\dot{\cX_0}'$ 
plays a similar role, and we also contrast the integer values $3$ and $6$ with 
the rational value $3.5$.  
\begin{figure}
\centering
\includegraphics[width= \linewidth]{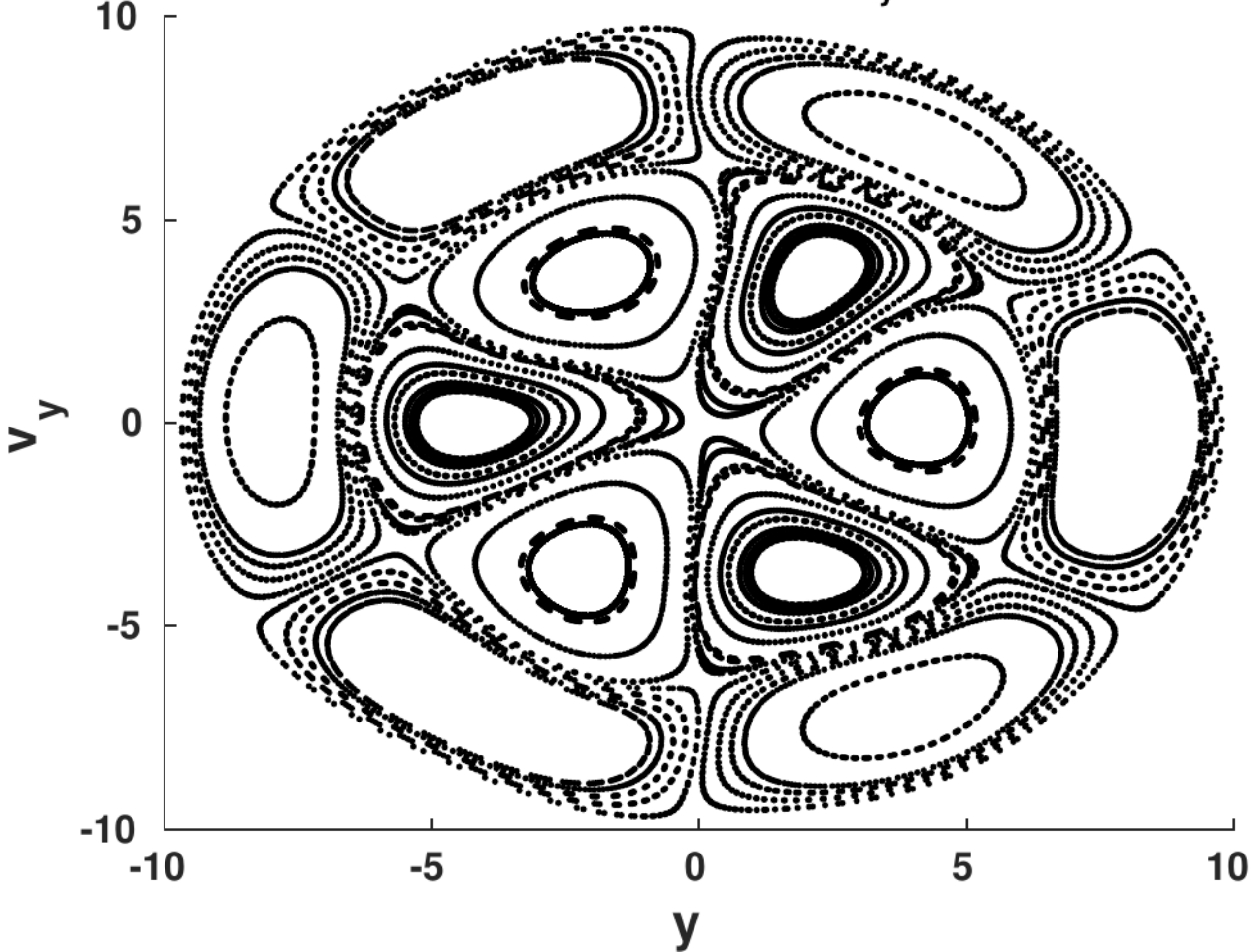}
\caption{Web network generated by Poincar\'e section of different trajectories 
with different initial ${\dot Y}'_0$ values and $X'_0=0$ for 
$\varepsilon = 0.5$, $\beta^2 = 1.75 \cdot 10^{-6}$ and $\nu_1 = 3$. 
%Due to small $\varepsilon <1$ value the Hamiltonian is integrable
}
\label{web_regular}
\end{figure}
\begin{figure}
\centering
\includegraphics[width= \linewidth]{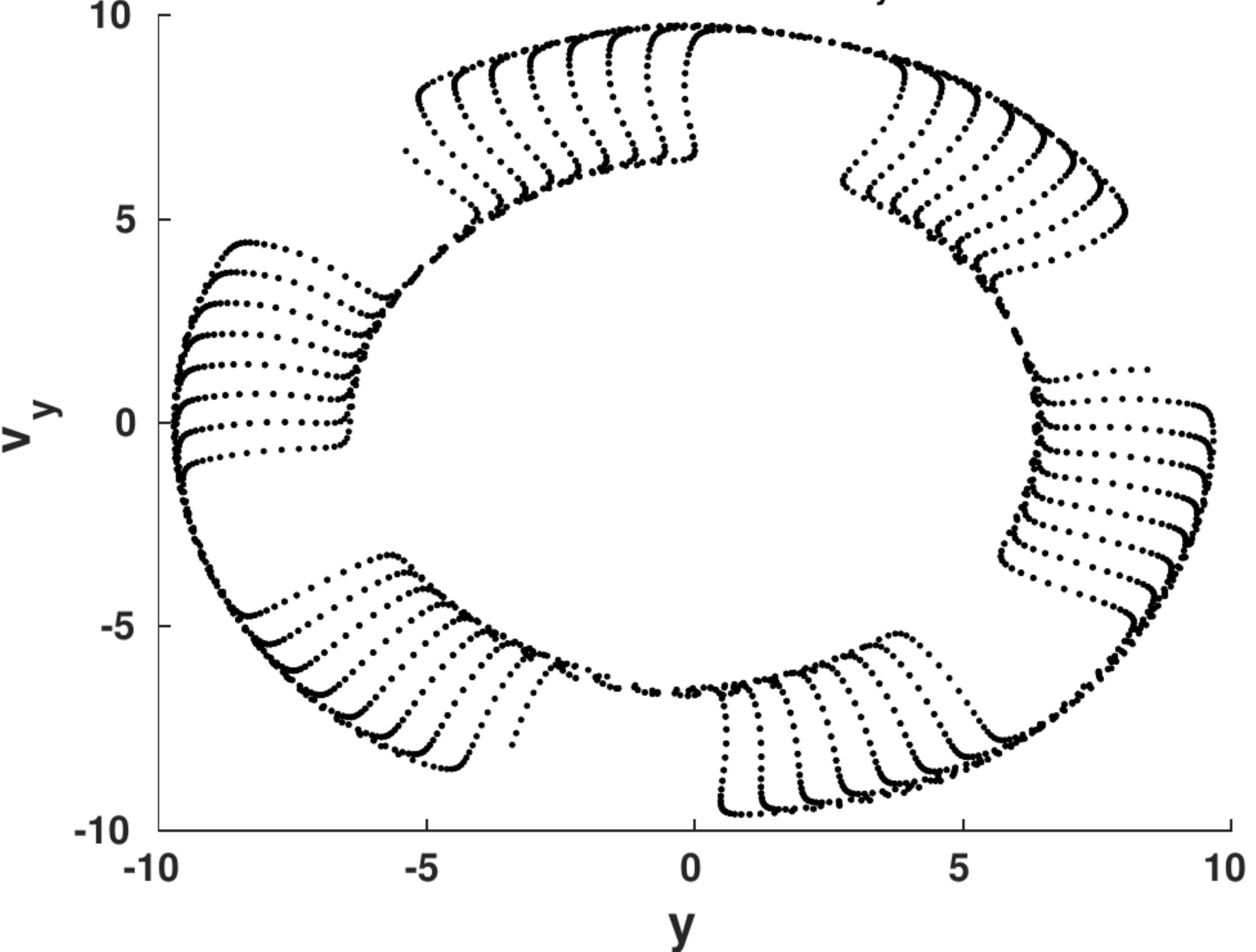}
\caption{Web network generated by Poincar\'e section of a single trajectory 
for $\varepsilon = 0.5$, $\beta^2 = 1.75 \cdot 10^{-4}$ and $\nu_1 = 3$. 
%Due to small $\varepsilon <1$ value the Hamiltonian is integrable. 
The rotation is due to fast change in $X'$.}
\label{web_regular_rotate}
\end{figure}

For small values $\varepsilon \beta^2 \ll 1$, 
the dynamics generate a spiral stochastic web \cite{vasilev91} in 
the three-dimension space ($X', Y', {\dot Y}'$). 
Moreover, for $\varepsilon \ll 1$, the dynamics is nearly integrable 
and the chaoticity vanishes. 
In the section cut by a plane $X'= \textrm{const}$, 
the web is a system of concentric circles and of straight lines 
passing through the co-ordinate origin in the ($Y', {\dot Y}'$) plane.
The cells of the web form concentric ``belts'' around the coordinate origin.
These trajectories are called ``web trajectories'', 
due to their special structural shape. 
Birkhoff's theorem ensures that, for small, nonzero $\varepsilon$ and $\beta$, 
the Poincar\'e map has O type (elliptic) fixed points and X type (hyperbolic) ones ;
regular trajectories are organized in islands around the O points, 
while chaotic ones wander through the heteroclinic and homoclinic 
tangles associated with the stable and unstable manifolds of the X points \cite{ElEs03}.

Fig.~\ref{web_regular} presents the web network for different initial particle positions, 
for $\varepsilon = 0.5$, $\beta^2 = 1.75 \cdot 10^{-6}$ and $\nu_1 = 3$. 
Due to small $\varepsilon$ value, the Hamiltonian is nearly integrable and KAM theory applies. 
When $X'$ varies, as a result of detuning from longitudinal cyclotron 
resonance ($k_x {\dot X} = \ell \omega_{\rm ce}$), 
the separatrix network and, consequently, the stochastic web 
rotate about the coordinate origin with angular velocity ${\dot X}' \sim \beta^2$.
But the shape and area of the shells remain constant, and a particle that is 
executing rapid rotation inside a cell sufficiently far from the separatrix of 
the averaged motion will never intersect the stochastic web and will move 
regularly. 

For high values of $\beta^2$ or ${\dot X_0}'$, $X'(t)$ varies rapidly
and the chaotic tangle network also rotates rapidly compared to the case with 
slowly varying $X'$. This precludes the presence of islands in the
$(Y', {\dot Y}')$ phase space for long time evolution. 
Fig.~\ref{web_regular_rotate} presents the rotation of a regular web trajectory
due to the high $\beta^2$ value.

In our simulations,  
we evolve the dynamics of 1024 particles having initial Gaussian velocity 
distribution with unit standard deviation along the $y$-direction and 
covariance $\langle {\dot X_0}' \, {\dot Y_0}' \rangle = 0$. Along the $x$-direction,  
we consider a very small standard deviation $\sigma_x = 0.001$, and choose
the square of wave vector ratio $\beta^2 \sim 10^{-6}$.
%%
%Since the web trajectory drifts rapidly for high value of ${\dot X_0}'$, 
%precluding the presence of islands,
%we consider a very small standard deviation $\sigma_x = 0.001$ 
%along $x$-direction.
%Hence, in our case, the time-dependent dynamics is more interesting 
%compare to the time-independent dynamics. 
We first consider the time-dependent dynamics in Sec.~\ref{sec:Web-time-dep}, 
then the time-independent cases in Sec.~\ref{sec:Web-time-indep}.

%%%%%%%%%%%%%%%%%%%%%%%%%%%%%%%%%%%%%%%%%
\subsection{Time-dependent Hamiltonian}
\label{sec:Web-time-dep}
%%%%%%%%%%%%%%%%%%%%%%%%%%%%%%%%%%%%%%%%%

For the time-dependent Hamiltonian, the dynamics depends on 
$(\varepsilon, \beta^2, \nu_1)$. Parameter $\varepsilon$ expresses the ratio 
of bounce frequency to cyclotron frequency, $\beta$ expresses the ratio 
of the parallel and perpendicular components of the wave electric field, 
and $\nu_1$ is the normalized frequency of the electrostatic wave in the drifted frame. 

A large value of $\beta^2$ or ${\dot X}'_0$ causes the dynamics detuning from the longitudinal 
resonance condition, $k_x {\dot X'} = \ell \omega_{\rm c}$, where $\ell$ is an integer. 
Therefore, the orbits in the web structure drift more rapidly away from their initial action values,  
covering the entire phase space inside the web and destroying islands.
In our present simulation, $\beta^2 \sim 10^{-6}$ and 
$\vert {\dot X'_0}\vert \ll 1 $ which induces a slow drift of the trajectory.

We evolve Eqs~(\ref{TimeDep_Red}) numerically for three different sets of 
parameters $(\varepsilon, \beta^2, \nu_1) = (3.21, 1.75 \cdot 10^{-6}, 6.0)$, 
$(3.21, 1.75 \cdot 10^{-6}, 3.0)$ and $(0.69, 1.83 \cdot 10^{-5}, 1.39)$, 
respectively. The frequency of the electrostatic mode $\omega_1$ is very 
small with respect to $\omega_{\rm c}$. In the drifted frame, one has $\nu_1 > \omega_{\rm c}$,
and the chaoticity criterion $\varepsilon > 0.25 \, \nu_1^{2/3}$
derived by Karney \cite{karney78} must hold. 
For all three sets of parameters, this stochasticity criterion is indeed satisfied.
\begin{figure}
\includegraphics[width=\linewidth]{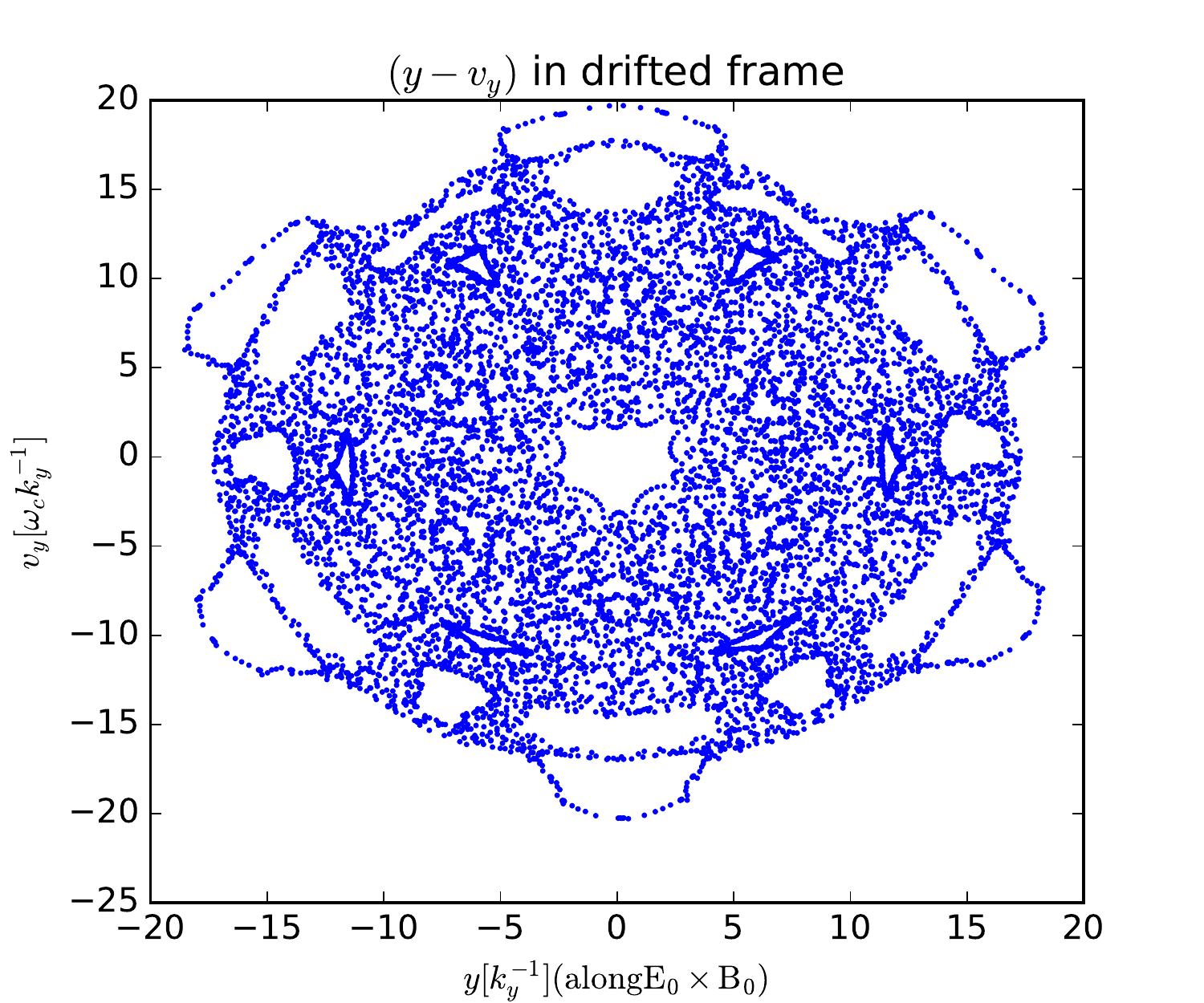}
\caption{Poincar\'e section of a single trajectory of (\ref{TimeDep_Red}) 
for $\varepsilon = 3.21$, $\beta^2 = 1.75 \cdot 10^{-6}$ 
and $\nu_1 = 6$.}
\label{web_6_fold}
\end{figure}
\begin{figure}
\includegraphics[width=\linewidth]{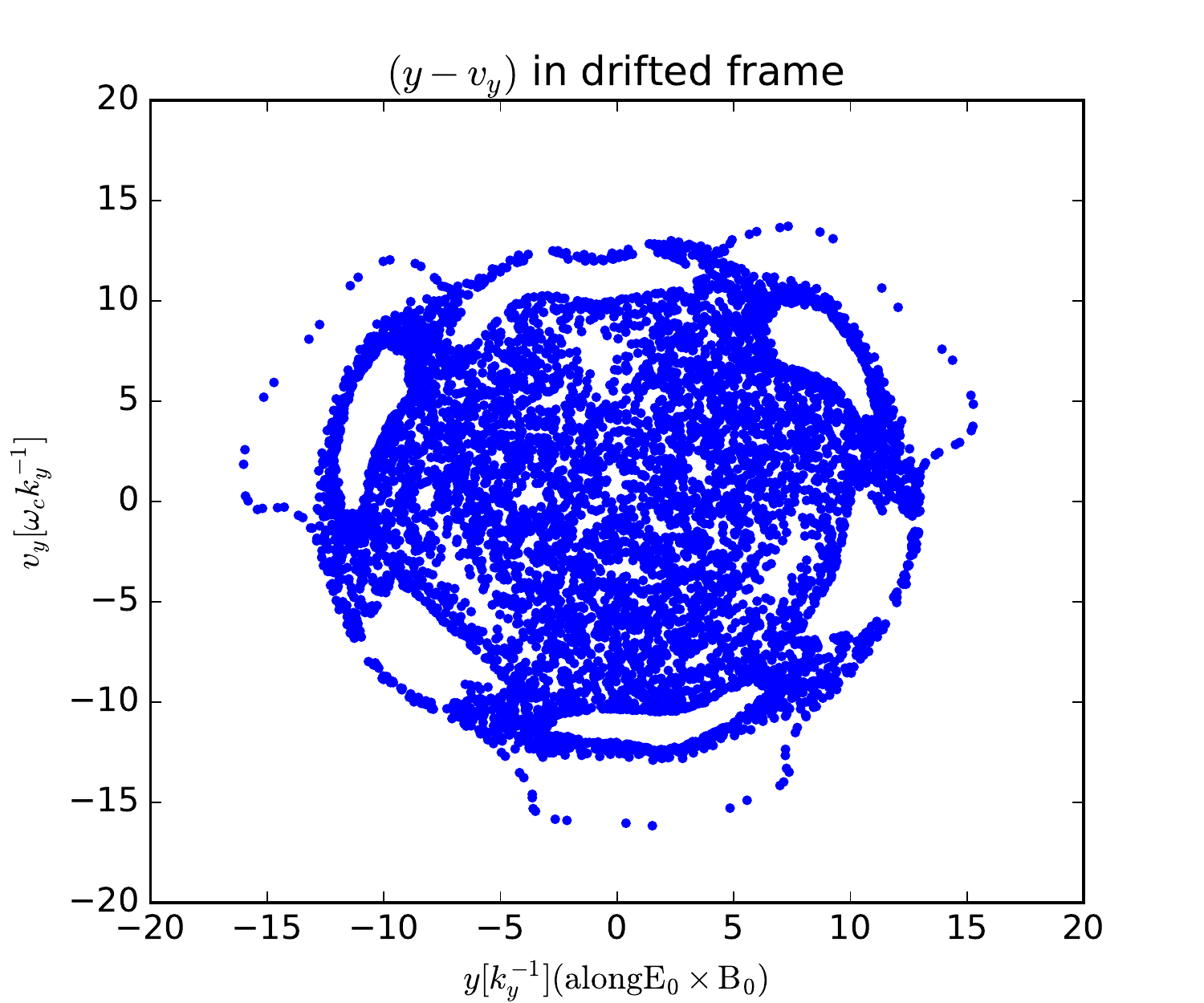}
\caption{Poincar\'e section of a single trajectory of (\ref{TimeDep_Red}) 
for  $\varepsilon = 3.21$, $\beta^2 = 1.75 \cdot 10^{-6}$ 
and $\nu_1 = 3$.}
\label{web_3_fold}
\end{figure}
\begin{figure}
\includegraphics[width=\linewidth]{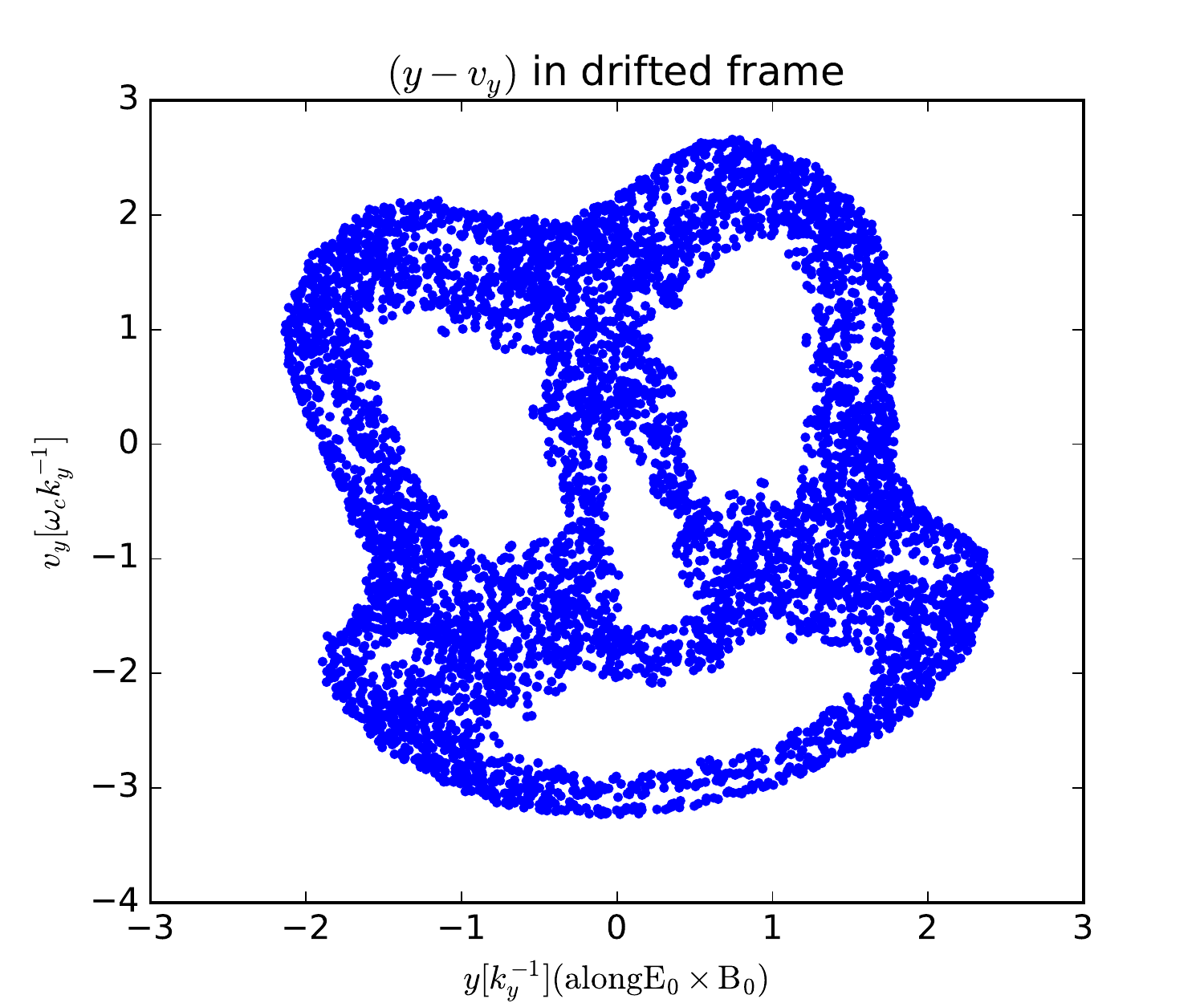}
\caption{Poincar\'e section of a single trajectory of (\ref{TimeDep_Red})  
for $\varepsilon = 0.69$, $\beta^2 = 1.83 \cdot 10^{-5}$  
and $\nu_1 = 1.39 $.}
\label{web_mask}
\end{figure}

We first plot the stroboscopic Poincar\'e section 
in the $(Y', P'_y)$ plane of a single particle trajectory 
at times $t = 2 n \pi / \nu_1$, where $n= 0, 1, 2...$. 
The parameter $\varepsilon$ 
determines the radius of the stochastic web. 
The value of $\nu_1$ determines the shape of the web structure. 
For integer $\nu_1$, we observe a web structure with $\nu_1$-fold rotational 
symmetry (Figs~\ref{web_6_fold} and \ref{web_3_fold}). 
For parameters 
$(\varepsilon, \beta^2, \nu_1) = (0.69, 1.83 \cdot 10^{-5}, 1.39)$
with non-integer $\nu_1$, 
the dynamics generates a Halloween mask-like, deformed three-fold web structure 
(Fig.~\ref{web_mask}).

Along $Y'$, the dynamics Eq.~(\ref{TimeDep_Red}) has two time scales, 
one associated with the electrostatic wave (with period $2 \pi / \nu_1$ in the drifting frame) 
and the other associated with a simple harmonic oscillator with period $2 \pi$.  
%(by making the right hand side zero of the $Y'$ component equation of 
%Eq.~(\ref{TimeDep_Red})). 
Therefore, an integer value of $\nu_1$ causes 
resonance between these two time scales and one can eliminate the time 
dependence by taking the Poincar\'e section at a regular time interval, 
$n T = 2 \pi n / \nu_1$, to generate the stochastic web 
structures in the Poincar\'e section plot. 
The reduced frequency $\nu_1$ will determine the shape of the web structure. 

For fixed values of $\varepsilon$, $\beta^2$ and $\nu_1$, any initial condition 
$(X'_0,Y'_0, {\dot X'}_0, {\dot Y'}_0)$, within the chaotic
domain of the stochastic web, generates a similar web structure, and the particles 
with initial conditions outside the web structure and well inside the sticky 
islands (regions with no points in the web structures) generate regular 
trajectories.

\begin{figure}
\includegraphics[width=\linewidth]{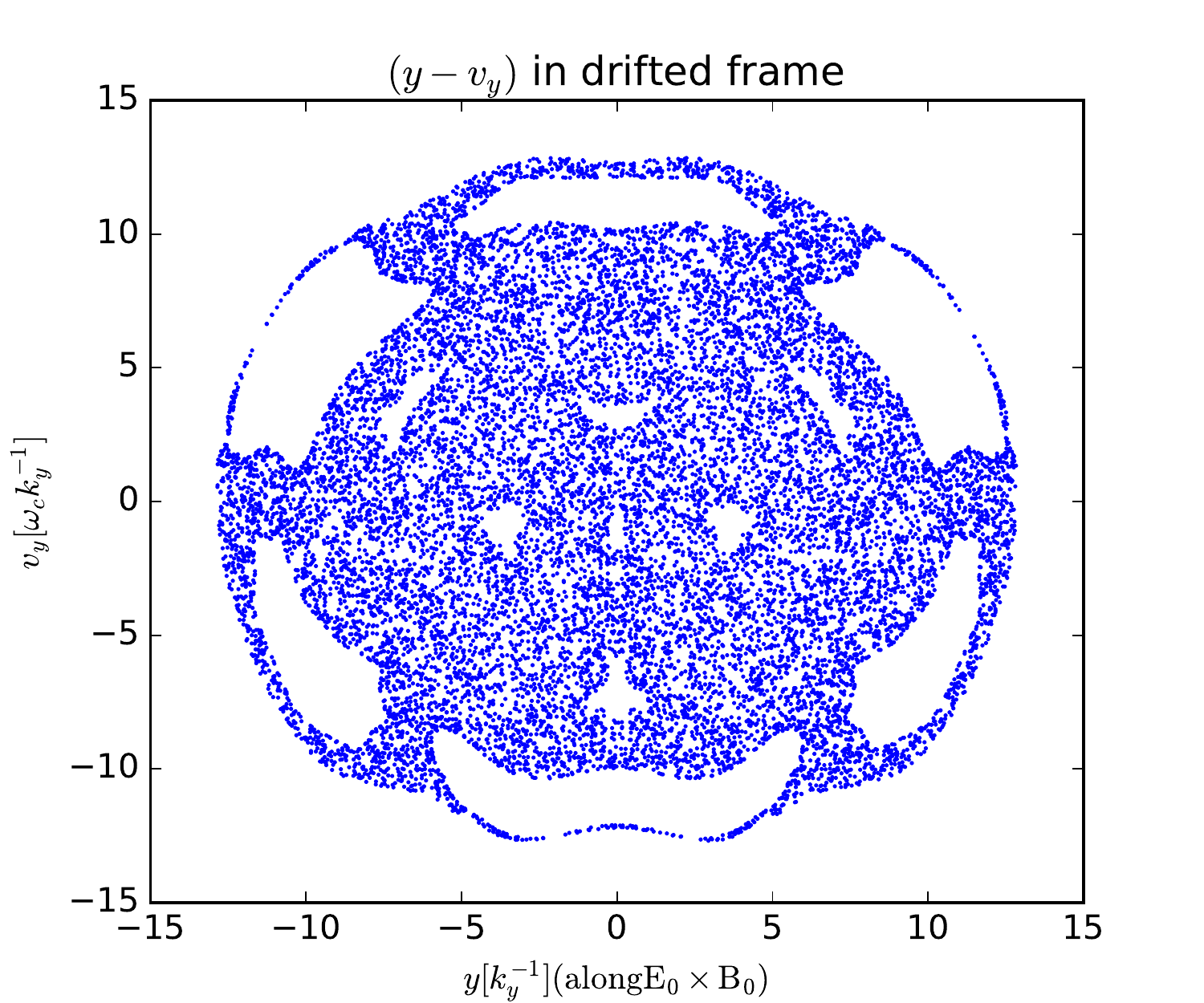}
\caption{Stochastic web of a single trajectory of (\ref{Auton_Red}) 
for $\varepsilon = 3.21$, $\beta^2 = 1.75 \cdot 10^{-6}$ 
and  ${\dot \cX_0}' = 3.0$.}
\label{web_vx_3_kx_0p001}
\end{figure}
\begin{figure}
\includegraphics[width=\linewidth]{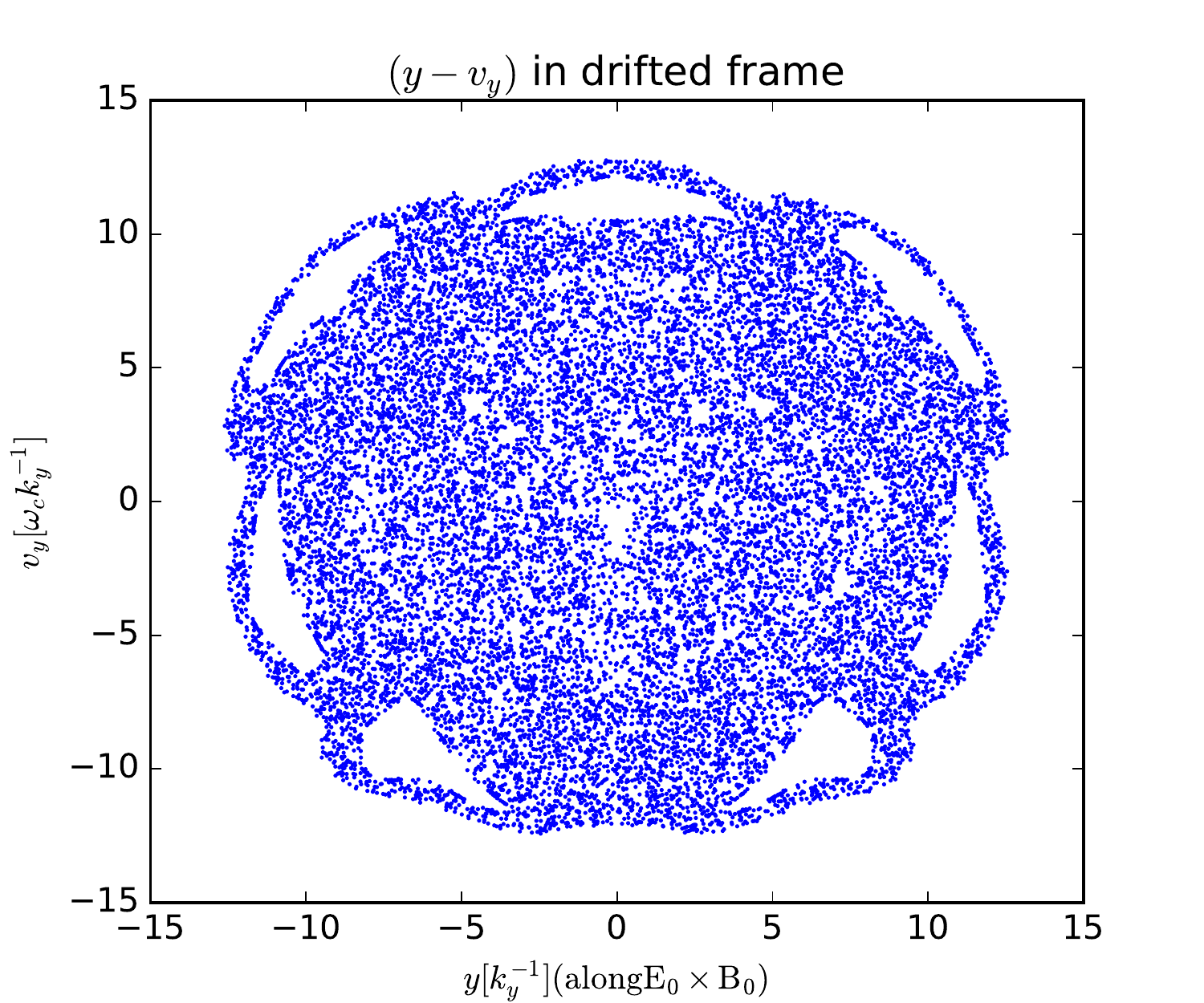}
\caption{Stochastic web of a single trajectory of (\ref{Auton_Red}) 
for $\varepsilon = 3.21$, $\beta^2 = 1.75 \cdot 10^{-6}$ 
and  ${\dot \cX_0}' = 7/2 = 3.5$.}
\label{web_vx_3p5}
\end{figure}
%
%%%%%%%%%%%%%%%%%%%%%%%%%%%%%%%%%%%%%%%%%
\subsection{Time-independent Hamiltonian}
\label{sec:Web-time-indep}
%%%%%%%%%%%%%%%%%%%%%%%%%%%%%%%%%%%%%%%%%
%
In the time-independent Hamiltonian, 
$\dot{\cX'} = {\dot X'} - \nu_1 $ and the dynamics depends on 
$(\varepsilon, \beta^2)$ only. In this case, for any initial condition 
$(\cX'_0,\cY'_0, \dot{\cX'}_0, \dot{\cY'}_0)$ within the chaotic
domain of the stochastic web, the shape of the web structure 
depends on the initial $\dot{ \cX'}_0$ values. 
For different $\dot{ \cX'}_0$ values, the trajectory lies on different energy surfaces  
$\cK = $~constant. Thus, particles with 
different initial conditions generate different web structures. Since 
$\beta^2 \sim 10^{-6}$ in this study, the motion along the $\cX'$ direction is 
almost ballistic, $\dot{\cX'} \cong \textrm{constant}$. Therefore, we can generate 
Poincar\'e section plots by taking sections at $\cX' = n 2\pi$, where $n$ is an 
integer. 

Figs.~\ref{web_vx_3_kx_0p001} and \ref{web_vx_3p5} 
display the stroboscopic plot of the time-independent 
dynamics Eq.~(\ref{Auton_Red}) with 
$(\varepsilon, \beta^{2}) = (3.21, 1.75 \cdot 10^{-6})$ 
and two different initial velocities along the $\cX'$ direction, 
$\dot{\cX}'_0 = 3$ and $3.5$ 
respectively. 
For integer values of $\dot{\cX'}_0$ as in Fig.~\ref{web_vx_3_kx_0p001}, 
the dynamics generates web structures 
similar to those generated in the Poincar\'e section plot for the cases of time-dependent 
dynamics~(\ref{TimeDep_Red}) with same integer values of $\nu_1$. 
Fig.~\ref{web_vx_3p5} presents the stroboscopic plot of a 
particle with $v_{0x} = 3.5$, which corresponds to a higher-order resonance ($7/2$)~: 
for fractional values of $\dot{\cX'}_0$, the 
stroboscopic plot generates different structures, because each of the 
different initial conditions lies on a different energy ($\cK = {\rm{constant}}$) surface. 
Therefore, the web structures in the time-independent dynamics 
highly depend on the initial conditions of the particle. 
%

%%%%%%%%%%%%%%%%%%%%%%%%%%%%%%%%%%%%%%%%
\section{Transport properties}
\label{sec:Transport}
%%%%%%%%%%%%%%%%%%%%%%%%%%%%%%%%%%%%%%%%
%
To characterise the transport properties, we consider a simple observable. 
Previous studies \cite{leoncini08, bouchara15} for time-dependent 
one-degree-of-freedom Hamiltonian systems 
focused on the norm of velocity $(\dot{p},\dot{q})$ in phase space, 
where $p,q$ are canonical co-ordinates. Here, we consider the arc length $s$ 
of the trajectory in position space only, or, in dimensionless variables of 
Eqs~(\ref{TimeDep_Red}), 
\begin{equation}
  S' (t) = \int_0^t \sqrt{\rmd X'^2 + \rmd Y'^2} .
  \label{arclengthXY}
\end{equation} 

Numerically, we consider the global average speed along the trajectory of a 
typical particle $i$
\begin{equation}
 \bar{v}_i(n) 
  = \frac{1}{n \Delta t} \sum_{k=0}^{n-1} 
    \sqrt{[ \Delta X'_i(t_k) ]^2 + [\Delta Y'_i(t_k)]^2}  \, ,
  \label{arc_length_XY_num}
\end{equation}
where $k$ is the timestep index, with coordinate increments
\begin{eqnarray}
  \Delta X'_i(t_k) 
  & = & X'_i(t_{k+1}) - X'_i(t_k)  \, ,
  \\
  \Delta Y'_i(t_k) 
  & = & Y'_i(t_{k+1}) - Y'_i(t_k)   \, .
\end{eqnarray}  
In eq.~(\ref{arc_length_XY_num}), the discretized form of the integral (\ref{arclengthXY}) is used.
\begin{figure}
\includegraphics[width=\linewidth]{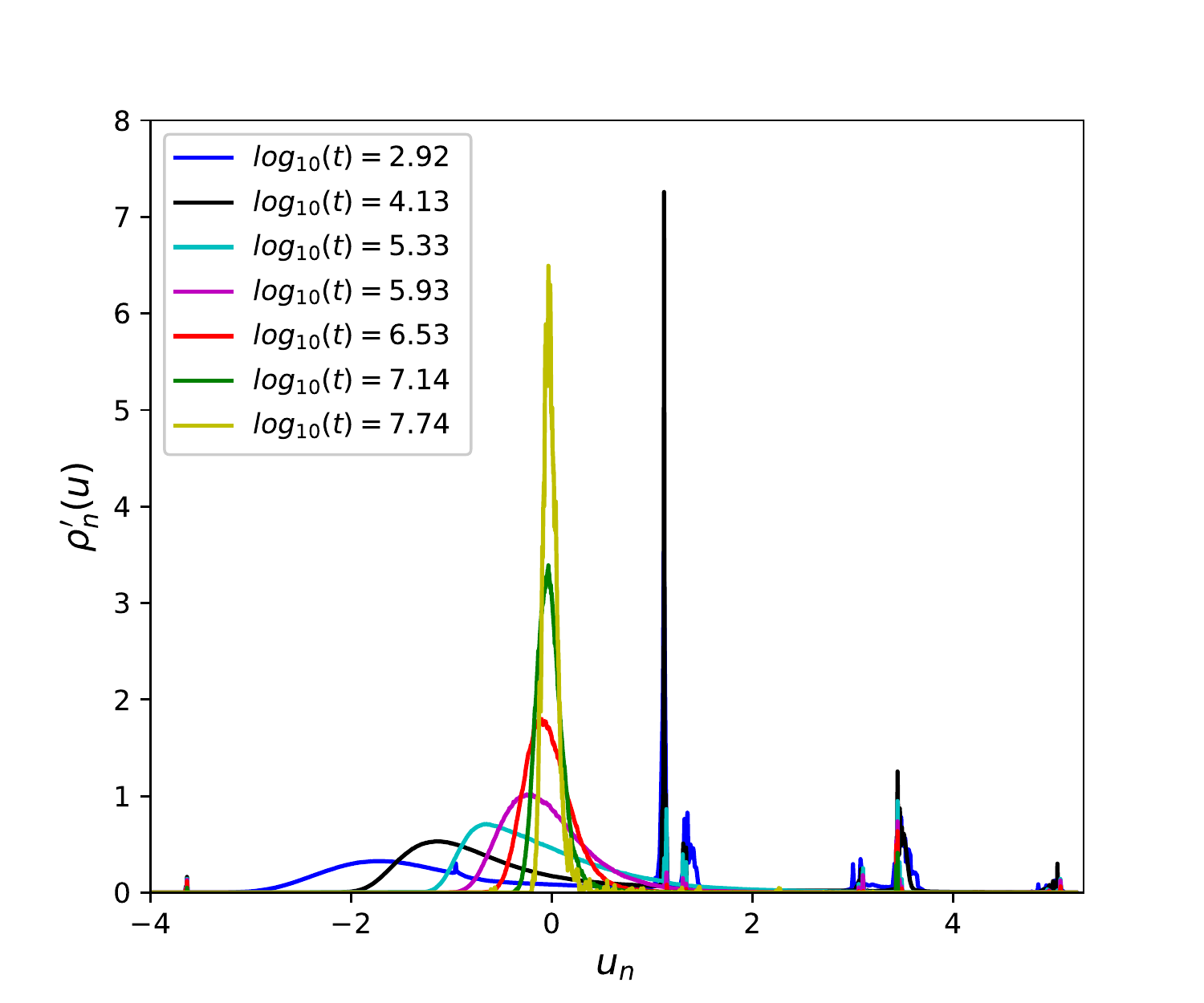}
\caption{Distribution of average speed for the stochastic web with $\nu_1 = 3$ at 
$\omega_{\rmpe} \, t = 8 \cdot 10^2, 13 \cdot 10^3, 21 \cdot 10^4, 85 
\cdot 10^4, 34 \cdot 10^5$, $13 \cdot 10^6$ and $5.4 \cdot 10^7$.}
\label{arc_distrib}
\end{figure}
\begin{figure}
\includegraphics[width=\linewidth]{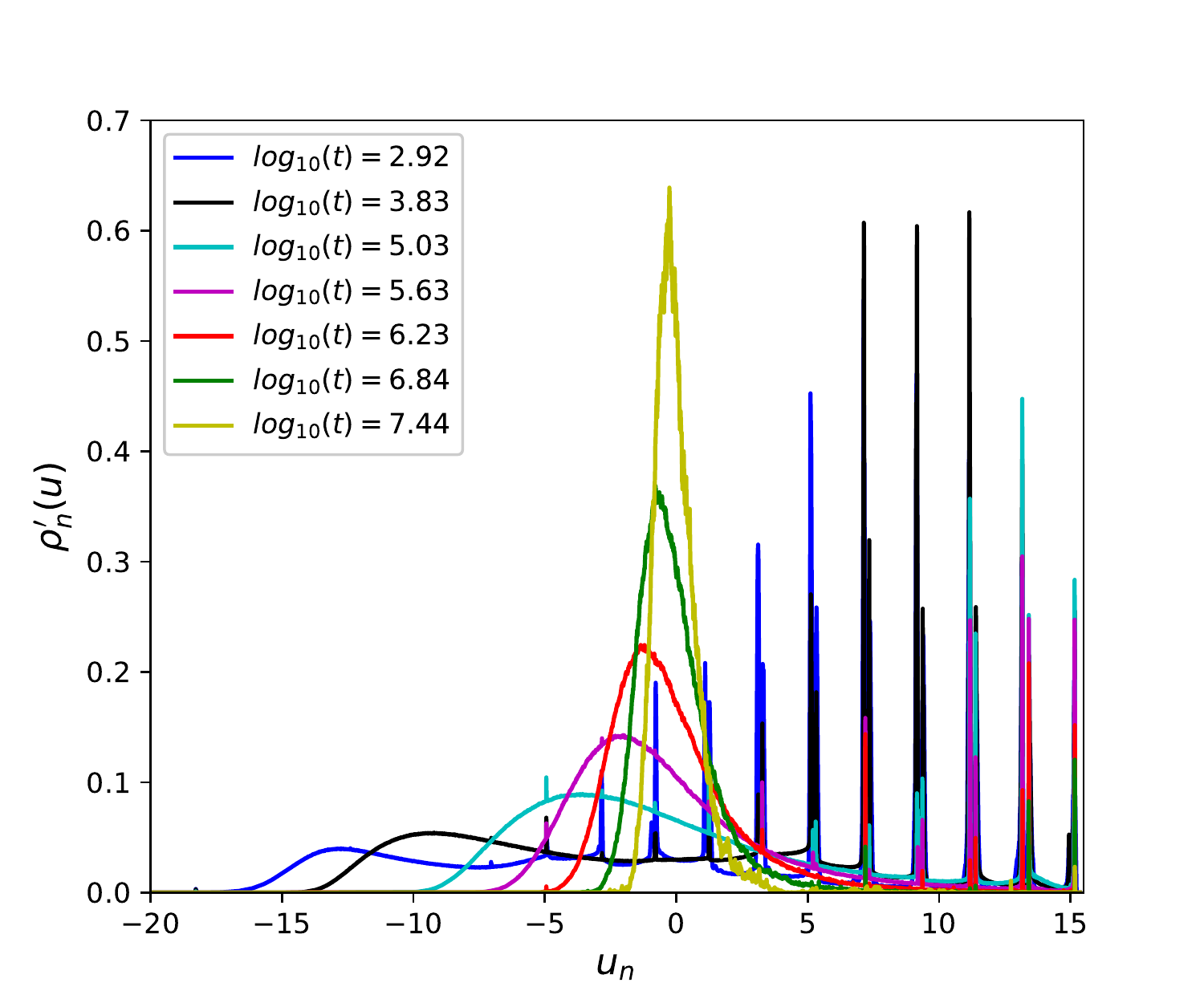}
\caption{
Distribution of average speed for the stochastic web with $\nu_1 = 6$ at 
$\omega_{\rmpe} \, t = 8 \cdot 10^2, 6.7 \cdot 10^3, 1.0 \cdot 10^5, 
4.2 \cdot 10^5, 1.7 \cdot 10^6$, $6.9 \cdot 10^6$ and $2.7 \cdot 10^7$.}
\label{arc_distrib_nu6}
\end{figure}

When conditions are met so that we can apply the central limit theorem, the distribution of quantity
\begin{eqnarray}\nonumber
\mathcal{Z}_i := \frac{1}{\sqrt{n}} \sum_{k=1}^n (v_i(k) - \langle v\rangle), 
\end{eqnarray}
gives a normal distribution function with mean $0$ and variance $\sigma^2$,
where $v_i(k) = \sqrt{[\Delta X'_i(t_k) ]^2 + [\Delta Y'_i(t_k)]^2}$
and $\langle v \rangle$ is defined below.
The difference between the quantity $\bar{v}_i$ and $\mathcal{Z}_i$ 
is the rescaling by $1/\sqrt{n}$ in $\bar{v}_i$. The variance of the 
distribution of $\bar{v}_i$ shrinks with rate $\sigma^2 \sim 1/\sqrt{n}$ as $n$ increases. 
Since the area under the distribution function is constant (equal to unity), 
the height of the distribution increases with rate $\sqrt{n}$ as the variance shrinks.
We define $\rho_n(\bar{v})$ as the sampling density of the distribution of $\bar{v}_i(n)$'s. 
Therefore, good ergodic properties of the dynamics~(\ref{TimeDep_Red}) would include the 
convergence of $\rho_n$ towards a Dirac distribution for $n \to \infty$, 
in which case the support $\langle v \rangle$ of the limit is the time 
average of the $\bar{v}_i(n)$'s, 
\begin{equation}
  \lim_{n \to \infty} \rho_n(\bar{v}) 
  = \delta(\bar{v} - \langle v \rangle),
  \label{arc_length_av}
\end{equation}
and, almost surely with respect to the initial condition (viz.\ index $i$),
\begin{equation}
  \langle v \rangle := \lim_{n \to \infty} \bar{v}_i(n)  .
  \label{eq:ergod_v}
\end{equation}
In the present case, what is understood by ergodic properties is that, 
when considering initial conditions in the chaotic sea, 
they will almost surely give rise to the same natural ergodic measure, 
meaning that they equally sample all the accessible chaotic domain of phase space 
when the dynamics evolve for sufficiently long time.
\begin{figure}
\includegraphics[width=\linewidth]{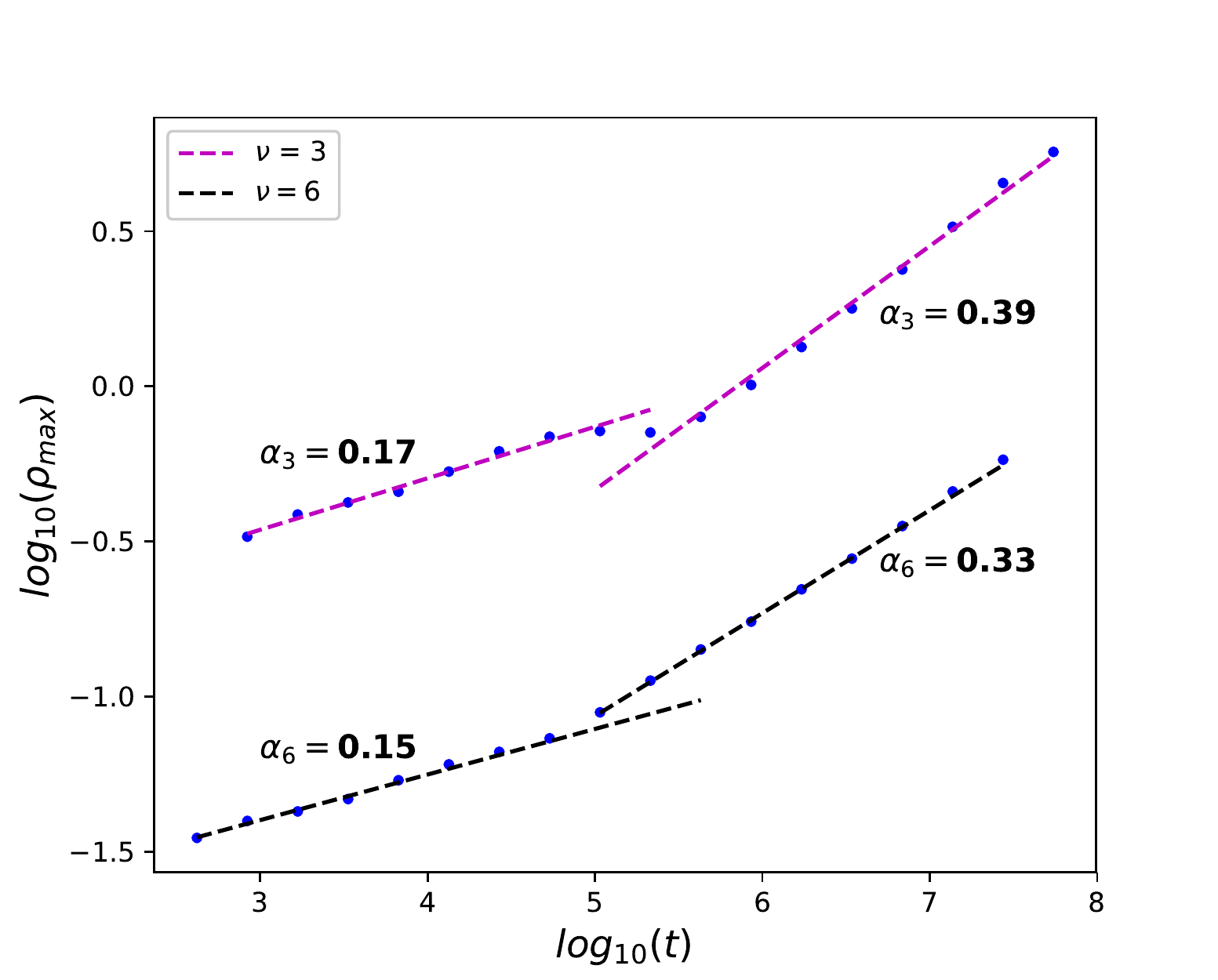}
\caption{Evolution of $\rho_{\max}$ versus $n$ for the time-dependent 
Hamiltonian with $\nu_1 = 3$ (magenta) and $\nu_1 = 6$ (black).}
\label{arc_max}
\end{figure}

One method to assess the convergence of $\rho_n$ is to look at how fast its 
maximum value $\rho_{\max}(n)$ diverges towards $+ \infty$ with $n$. 
%When the dynamics is sufficiently chaotic in the sense that 
In order to characterize this speed, one can typically expect that a scaling applies to increments in (\ref{arc_length_XY_num}),  
so that one may expect 
\begin{equation}
  \rho_{\max}(n) \sim n^{\alpha} ,
  \label{arc_length_max}
\end{equation}
where the exponent $\alpha$ characterises the nature of the transport.
If increments in (\ref{arc_length_XY_num}) are quite independent 
and a central limit theorem applies,
transport is diffusive and $\alpha = 1/2$. 
For $\alpha > 1/2$ it is sub-diffusive,
 and for $\alpha < 1/2$ it is super-diffusive.

Indeed, instead of considering the global average quantity (speed) $\bar{v}_i$, 
one may consider the arc length  
%\begin{eqnarray}\nonumber
$	S_i(t) = \sum_{k=1}^n 
        \sqrt{[\Delta X_i'(t_k)]^2+[\Delta Y_i'(t_k)]^2} = n \bar{v}_i$
%\end{eqnarray}
in order to characterise the transport. 
Here, when we can apply
%Therefore applying
the central limit theorem,
one expects the variance of the distribution of $S_i$'s to grow like $\sqrt{n}$.  
One can also calculate different moments $\mathcal{M}_{\mathfrak{q}}$ 
of the distribution of $S_i$'s and 
extract the characteristic exponent $\mu$ from
\begin{eqnarray}%\nonumber
   \mathcal{M}_{\mathfrak {q}} 
   = 
   \langle \vert S_i - \langle S_i\rangle\vert^{\mathfrak{q}}\rangle 
   \sim 
   t^{\mu(\mathfrak{q})}.
\end{eqnarray}
The second order exponent $\mathfrak{q}=2$ may be related \cite{bouchara15} 
to the variance of the displacement in space, $\sigma^2 \sim t^{\mu(2)}$. 
For a chaotic system of diffusive type, $\mu(2) = 1$, 
while $\mu(2) \ne 1$ for an anomalous transport.

Moreover, the area under the distribution function is constant, which implies
that the variance grows faster in presence of fat tails than in the diffusive case. 
Therefore, $\mu(2)> 1$ implies super-diffusive transport 
and $\mu(2)<1$ implies sub-diffusive transport. 
The exponent $\alpha$ associated with $\rho_{\rm max}$ of the speed distribution 
is different from this $\mu(2)$ which is associated with the 
moment of the position-displacement distribution, 
but in the same spirit and due to the fact that normalized distributions have constant area, 
the existence  of fat tails will imply that the maximum $\rho_{\rm max}$ will not grow so fast, 
and thus $\alpha < 1/2$ for $\rho_{\rm max}$ indicates super-diffusive transport,
whereas $\alpha > 1/2$ indicates sub-diffusive transport. 
Both the scaling parameter $\alpha$ and $\mu(2)$ characterise the diffusion property,
and one may derive \cite{leoncini08} a relation $\alpha = 1 - \mu(2)/2$ between them.

%Since in presence of fat tail the maximum of
%the distribution increases slower than the diffusive case, therefore 
%$\alpha < 1/2$ in case of $\rho_{\rm max}$ calculation indicates the existence
%of super-diffusive transport and $\alpha > 1/2$ for sub-diffusive transport.

The non-diffusive aspect of transport implies that the usual central limit theorem does not apply. 
However, the motion of trajectories in the chaotic web does lose correlations, and 
each trajectory typically generates a similar picture of the web. 
Thus, one can still view a trajectory as a sequence of pieces which are essentially mutually independent,
but with duration in time and extent in space which do not meet 
the finite variance assumption necessary for the gaussian central limit. 
In other words, the successive pieces generate a process obeying a scaling law, 
viz.\ it is infinitely divisible, 
and their (properly rescaled) sum has a L\'evy distribution 
characterized by the scaling exponent \cite{balescu97,zaslavsky07}.

%%%%%%%%%%%%%%%%%%%%%%%%%%%%%%%%%%%%%%%%%%%%%%%%%%%%%%%%%%%
\subsection{Time-dependent Hamiltonian}
\label{sec:time-dep-transport}
%%%%%%%%%%%%%%%%%%%%%%%%%%%%%%%%%%%%%%%%%%%%%%%%%%%%%%%%%%%

In Figs.~\ref{arc_distrib} and \ref{arc_distrib_nu6}, we plot the 
distribution of 
\begin{equation}
  u_{i,n} = \bar{v}_i(n) - \langle v \rangle
  \label{eq:un}
\end{equation}
for two different web structures, with $\nu_1= 3$ and $\nu_1= 6$, respectively.
One can calculate the arc length for a time-independent dynamics also,  
but in the time-dependent dynamics, the parameter 
$\nu_1 = (v_{\rm d}k_y - \omega_1)/\omega_c $ is important in Hall thrusters. 
It expresses the frequency of the electrostatic modes generated 
by the $\bfE \times \bfB$ instability, 
in a frame which moves with the drift velocity $v_{\rm d}$ 
along the $\bfE_0 \times \bfB_0$ direction. 
%In Hall thrusters, $\nu_1$ is typically close to $3$. 
As our study is motivated by the anomalous chaotic transport in Hall thruster 
devices, we choose the time-dependent dynamics for characterising the 
transport for $\nu_1 = 3$ and $\nu_1 = 6$.

To characterize the transport, 
we consider two different stochastic webs 
corresponding to values 
$(3.21, 1.75 \cdot 10^{-6}, 3)$ and 
$(3.21, 1.75 \cdot 10^{-6}, 6)$
for parameters $(\varepsilon, \beta^2, \nu_1)$.
We evolve the equations of motion~(\ref{TimeDep_Red}) for 1024 particles 
with all initial conditions inside the chaotic domain,  
we calculate the arc length for each particle trajectory for a long time evolution 
($10^8 \, \omega_{\rmpe}^{-1}$) using a timestep value 
$\Delta t = 3.33 \cdot 10^{-3} \, \omega_{\rmc}^{-1}
 = 3.33 \cdot 10^{-2} \, \omega_{\rm{pe}}^{-1}$ in simulation,   
and we generate the distribution of the global average speed. 

We plot the distribution at different times in 
Figs.~\ref{arc_distrib} and \ref{arc_distrib_nu6} for both cases.
To avoid nonphysical peaks in the distribution of $\rho_{n}$, 
the length $n$ of the time sequence should be sufficiently long 
for the dynamics to reach a saturation state, 
i.e.\  for the Poincar\'e section of each particle's trajectory to sample the 
whole phase space reach of the web. 
Here, we construct the distribution functions $\rho_{n}$ 
at times $t_n \geq  800 \, \omega_{\rmpe}^{-1}$. 

In the plot, the strong sharp peaks are associated with the stickiness 
phenomenon, by which a trajectory may remain for a long time close to the 
regular islands. 
The number of sharp peaks depends on the number of resonance 
generating sticky islands within the web structures, 
which we further discuss in the next section.  
We obtain more peaks for $\nu_1 = 6$ (Fig.~\ref{arc_distrib_nu6}) 
than for $\nu_1 = 3$ (Fig.~\ref{arc_distrib}). 
In the figure, $\log_{10} t = \log_{10} (n \tau_{\mathrm P})$, 
where $\tau_{\mathrm P}$ is the average time between two successive 
Poincar\'e sections. The relative magnitude of these sharp peaks decreases 
as $n \rightarrow \infty$,  
because the contribution from the chaotic
domain becomes large compared to the contribution from the sticky regular 
trajectories as we consider a longer time evolution. 

In both Figs.~\ref{arc_distrib} and \ref{arc_distrib_nu6}, the 
distribution after time $t_n \sim 10^7 \, \omega_{\rmpe}^{-1}$ (yellow line) 
has almost zero relative strength of the sharp peaks, compared to the height 
of the smoother distribution. 
Stickiness generates a memory effect and L\'evy flights \cite{leoncini04}. 
In absence of these sticky trajectories, 
the transport is purely diffusive and the exponent $\alpha$ takes the value 
$1/2$. In the presence of these sticky trajectories, 
the transport will be anomalous. 

To measure $\alpha$, we find the value of $\rho_{\max}$ from 
the local maximum of the central smooth flat peak location 
which corresponds to the bulk and particles evolving in the chaotic domain, 
and not the sticky domains. 
In Fig.~\ref{arc_max}, we plot the time evolution of $\rho_{\max}$ 
for both cases $\nu_1 = 3$ and $\nu_1 = 6$. From the 
curve fitting, we obtain two different values $\alpha_3 = 0.17$ and 
$0.39$ for the case $\nu_1 = 3$ (magenta dashed line and data dots), and 
$\alpha_6 = 0.15$ and $0.33$ for $\nu_1 = 6$ (black). 
Both values of $\alpha$ in both cases are below $0.5$. 
Thus, the diffusion is anomalous and super-diffusive. 

After a longer time evolution, most of the particles spend more time exploring 
the chaotic region 
%within the chaotic region 
of the stochastic web and sampling the ergodic measure. 
%Then, 
It appears that the contribution from the sticky islands 
``decreases'' in comparison with the contribution from the chaotic domain,
%Therefore 
and that the ``diffusion'' rate increases at longer time 
($t_n > 10^5 \, \omega_{\rmpe}^{-1}$).   
Note however that, even for large times, 
the exponent being smaller than 1/2 implies that 
the average speed fluctuations $\bar v_i - \langle v \rangle$ do not 
approach a Gaussian distribution, 
hence they do not obey the central limit theorem over this time scale, 
and transport is superdiffusive.

%%%%%%%%%%%%%%%%%%%%%%%%%%%%%%%%%%%%%%%%%%%%%%%%%%%%%%%%%%%
\subsection{Time-independent Hamiltonian}
\label{sec:Transport-time-indep}
%%%%%%%%%%%%%%%%%%%%%%%%%%%%%%%%%%%%%%%%%%%%%%%%%%%%%%%%%%%

Similarly, we analyse transport for a stochastic web structure 
generated from the time-independent Hamiltonian with the corresponding arc length 
\begin{equation}
  \cS' (t) = \int_0^t \sqrt{\rmd \cX'^2 + \rmd \cY'^2} .
  \label{arclengthXY-t-ind}
\end{equation} 

{\it  Transport in web with simple rational parameter}:
The trajectories of 1024 particles are computed numerically up to time $10^8 \, \omega_{\rmpe}^{-1}$ 
with time step $\Delta t = 5.5 \cdot 10^{-3}$. 
All particles are initially randomly distributed in $(\cX', \cY')$ plane 
within $-\pi \leq \cX'_{0} \leq \pi$ and $-\pi \leq \cY'_{0} \leq \pi$. 
Their initial velocities along the $\cY'$-direction
are drawn from a Gaussian distribution with unit standard deviation.
Along the $\cX'$-direction, we consider three different values 
$\dot{\cX}'_0 = 3$, $3.5$ and $6$ (simple rational) in order to analyse the 
transport in three different web structures generated 
from the time-independent Hamiltonian~(\ref{Auton_Red}).   
For all three cases, we consider $\varepsilon = 3.21$ and 
$\beta^2 = 1.75 \cdot 10^{-6}$. 
\begin{figure}
\includegraphics[width=\linewidth]{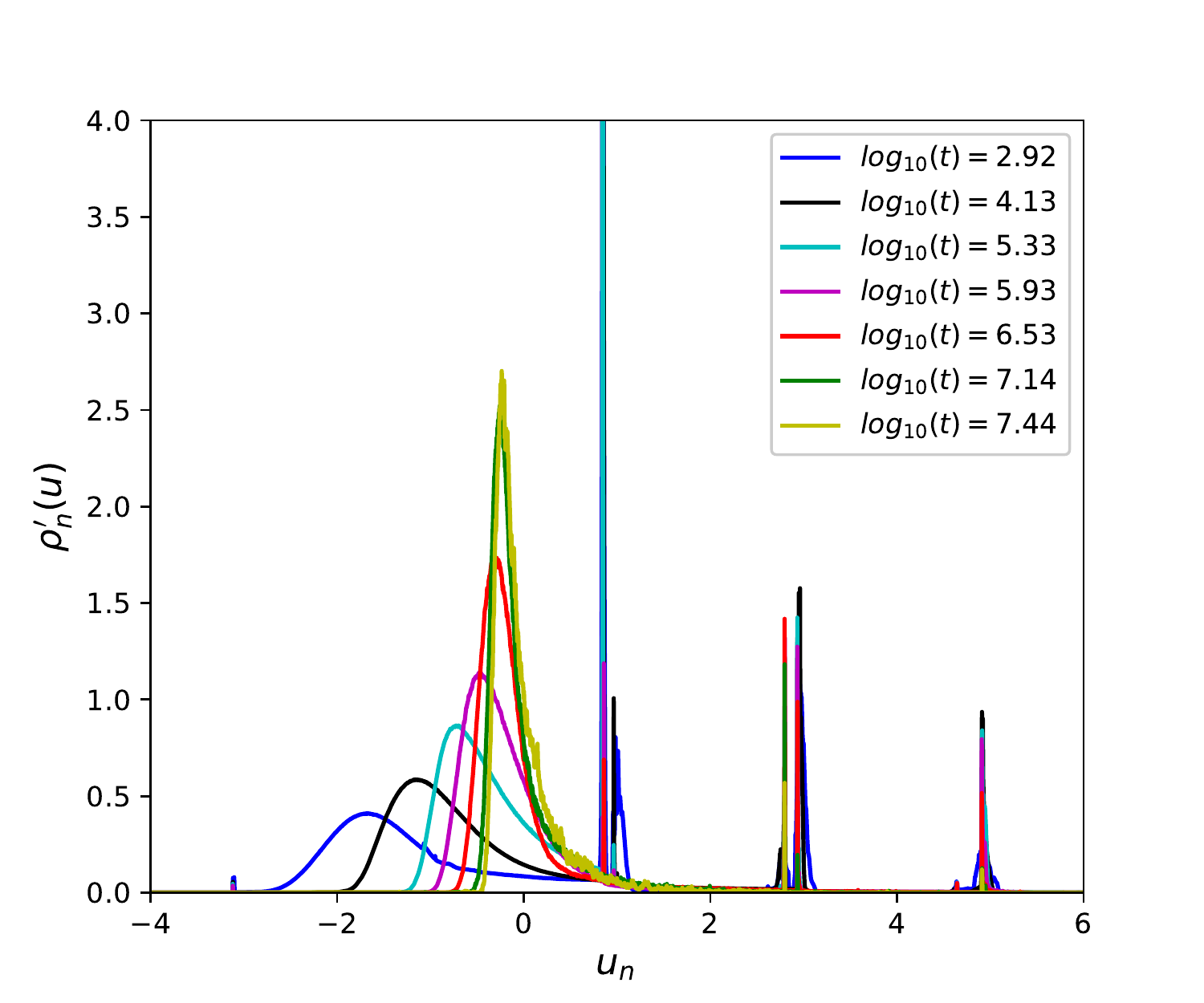}
\caption{Distribution of average speed for the stochastic web with 
$\dot{\cX}'_0 = 3$ and $(\varepsilon, \beta^2)= (3.21, 1.75 \cdot 10^{-6})$ at 
$\omega_{\rmpe} \, t = 8 \cdot 10^2$, $13 \cdot 10^3, 21 \cdot 10^4, 85 \cdot 10^4, 
34 \cdot 10^5$, $13 \cdot 10^6$ and $5.4 \cdot 10^7$.}
\label{arc_distrib_v0x_3}
\end{figure}
\begin{figure}
\includegraphics[width=\linewidth]{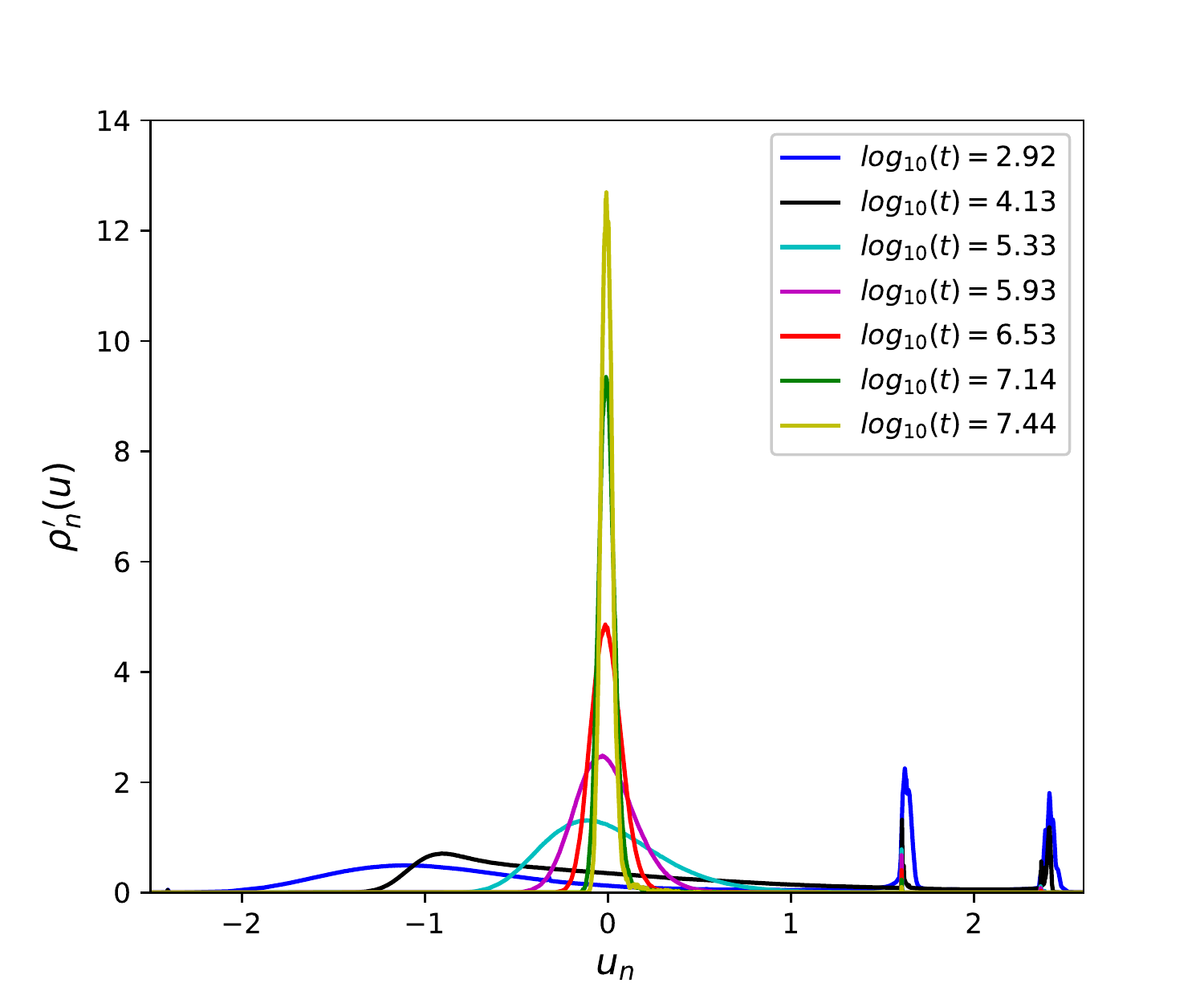}
\caption{ Distribution of average speed for the stochastic web with 
$\dot{\cX}'_0 = 3.5$ and $(\varepsilon, \beta^2)= (3.21, 1.75 \cdot 10^{-6})$ at 
$\omega_{\rmpe} \, t = 8 \cdot 10^2$, $13 \cdot 10^3, 21 \cdot 10^4, 85 \cdot 10^4, 
34 \cdot 10^5$, $13 \cdot 10^6$ and $5.4 \cdot 10^7$.}
\label{arc_distrib_v0x_3p5}
\end{figure}
\begin{figure}[t]
\includegraphics[width=\linewidth]{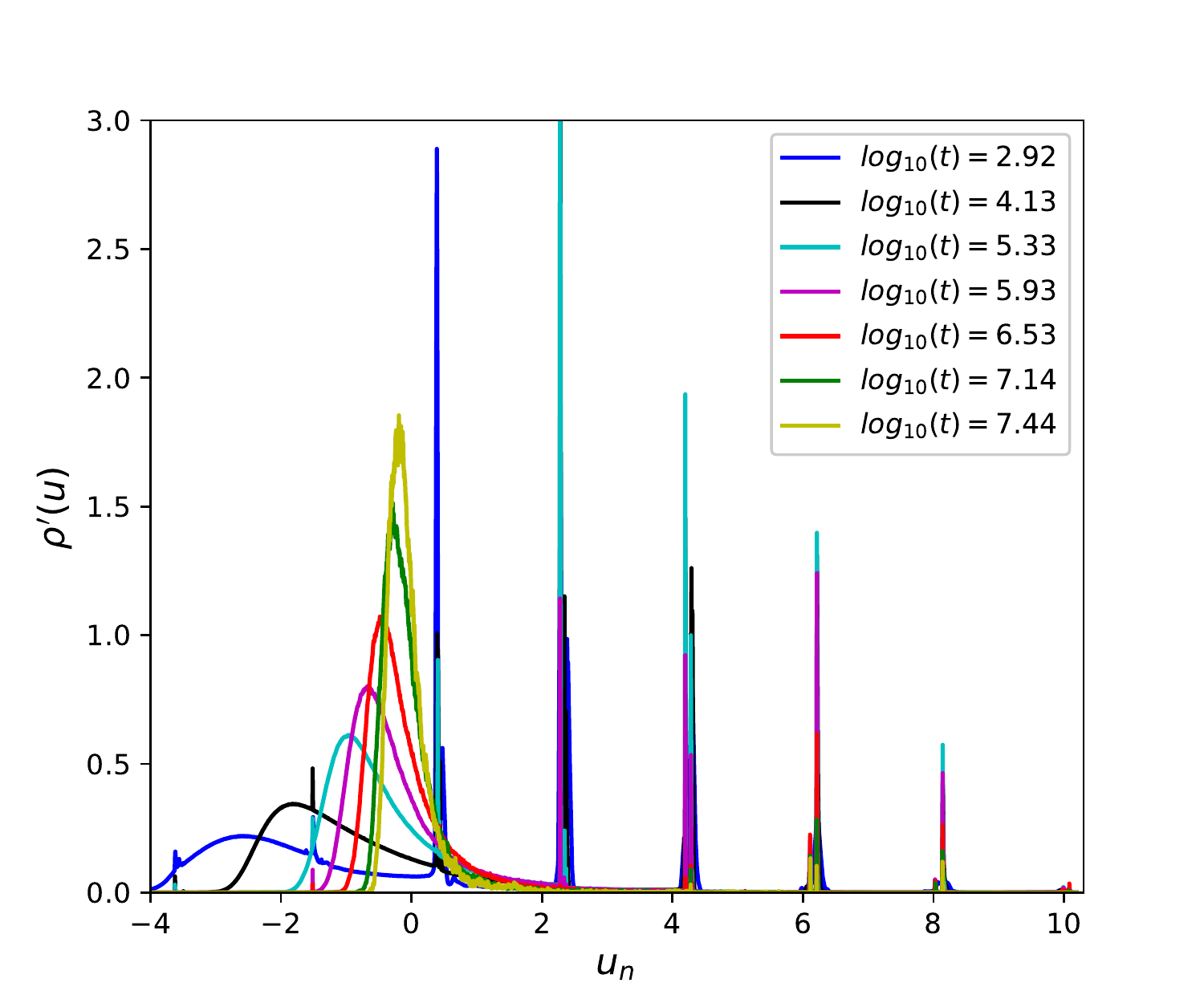}
\caption{ Distribution of average speed for the stochastic web with 
$\dot{\cX}'_0 = 6$ and $(\varepsilon, \beta^2)= (3.21, 1.75 \cdot 10^{-6})$ at 
$\omega_{\rmpe} \, t = 8 \cdot 10^2$, $6.7 \cdot 10^3, 1.0 \cdot 10^5, 4.2 \cdot 10^5, 
1.7 \cdot 10^6$, $6.9 \cdot 10^6$ and $2.7 \cdot 10^7$.}
\label{arc_distrib_v0x_6}
\end{figure}

Figs.~\ref{arc_distrib_v0x_3}, \ref{arc_distrib_v0x_3p5} and 
\ref{arc_distrib_v0x_6} present the distribution of the average speed at 
different times for all three cases. 
Similarly to the time-dependent cases, 
sharp peaks in the distribution of average speed appear 
due to the presence of the sticky islands, 
and the number of sharp peaks is larger 
for the web structures with six-fold rotational symmetry, $\dot{\cX'_0}= 6$, 
than for the three-fold rotational symmetry, $\dot{\cX'_0}= 3$. 
In the case $\dot{\cX'_0}= 3.5$, as seen in Fig.~\ref{web_vx_3p5}, 
the number and area of sticky islands 
are smaller than for $\dot{\cX'_0}= 3$.
Therefore, the height in the smooth part of the distribution, due to the chaotic 
domain, is larger for $\dot{\cX'_0}= 3.5$.  
\begin{figure}[t]
\includegraphics[width=\linewidth]{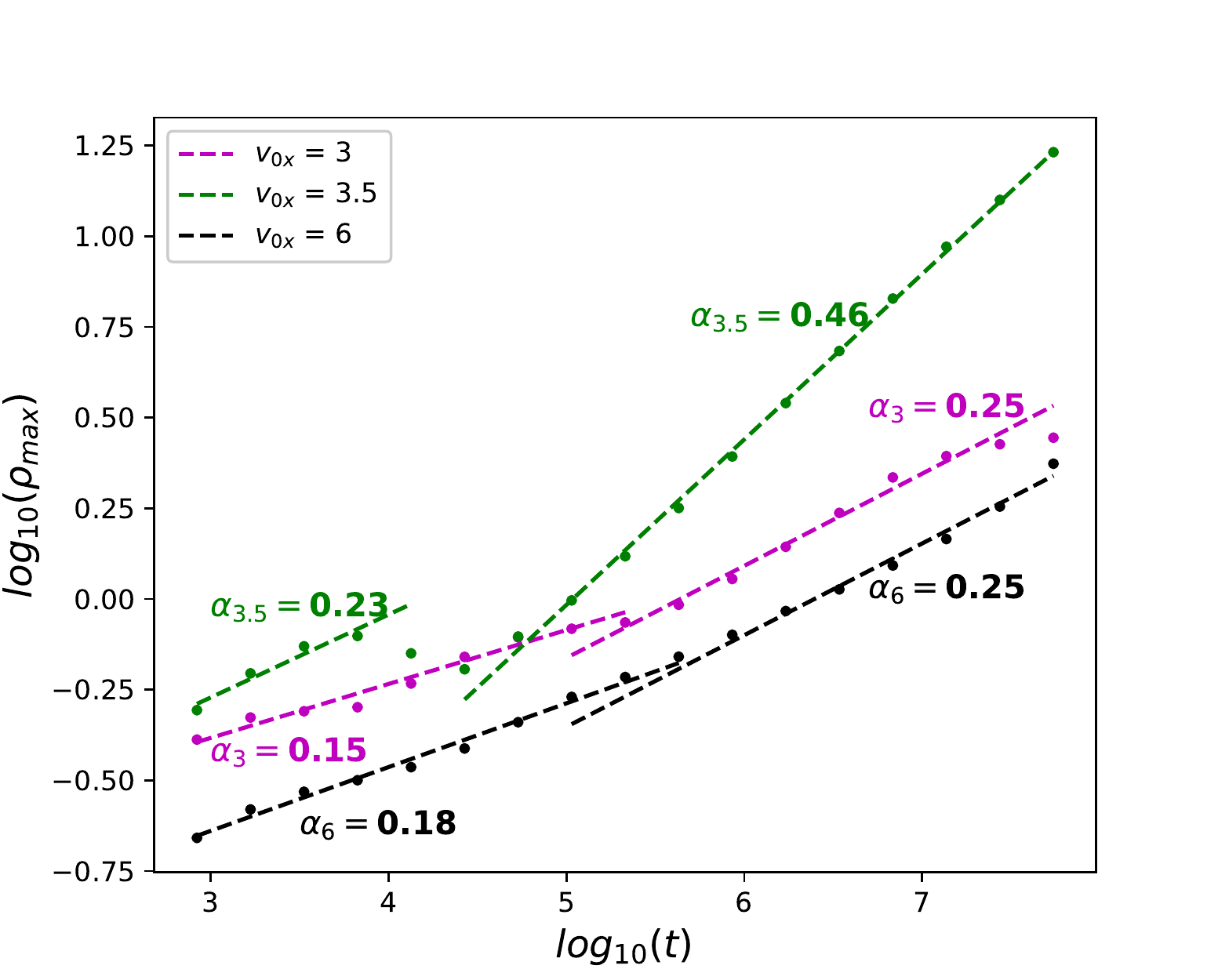}
\caption{Evolution of $\rho_{\max}$ versus $n$ for the time-independent 
Hamiltonian for $(\varepsilon, \beta^2)= (3.21, 1.75 \cdot 10^{-6})$ 
with $\dot{\cX}'_0 = 3$ (magenta), $\dot{\cX}'_0 = 6$ 
(black) and $\dot{\cX}'_0 = 3.5$ (green).}
\label{arc_max_tind}
\end{figure}

To estimate the exponent values from Eq.~(\ref{arc_length_max}), 
we plot in Fig.~\ref{arc_max_tind} the time evolution of $\rho_{\max}$ for all three cases. 
For $\dot{\cX}'_0 = 3$ (magenta) and $6$ (black), 
the plots are similar to the time-dependent cases. 
From the curve fitting, we obtain two different values of $\alpha$ in two 
different regimes of the plots, $\alpha_3 = (0.15, 0.25)$ and 
$\alpha_6 = (0.18, 0.25)$. 
In both cases, the transport is anomalous of super-diffusive type. 
The values for the shorter time span ($t < 5 \cdot 10^5 \, \omega_{\rmpe}^{-1}$) 
are very close to the values that are recovered for the time-dependent dynamics.

For fractional values of $\dot{\cX}'_0$, the number and area of the sticky 
islands are smaller than for the other two cases. 
Most of the region within the stochastic webs is part of the chaotic domain, 
therefore the height of the distribution increases at a higher rate 
than in the other two cases, as we increase the value of $n$. 
For $\dot{\cX}'_0 = 3.5 = 7/2$, we find higher exponent values, $\alpha_{3.5} = (0.23, 0.46)$. 
At short time, the relative contribution from sticky trajectories is significantly large, 
which reduces the exponent value to $\alpha_{3.5} = 0.23$~; 
in contrast, for longer time $t > 5 \cdot 10^4 \, \omega_{\rmpe}^{-1}$, 
the contribution from the chaotic region dominates over the 
contribution from sticky islands, and sharp peaks almost 
disappear, which increases the exponent value to $\alpha_{3.5} = 0.46$, 
so that the transport becomes closer to a diffusive type. 

\begin{figure}[t]
\includegraphics[width=\linewidth]{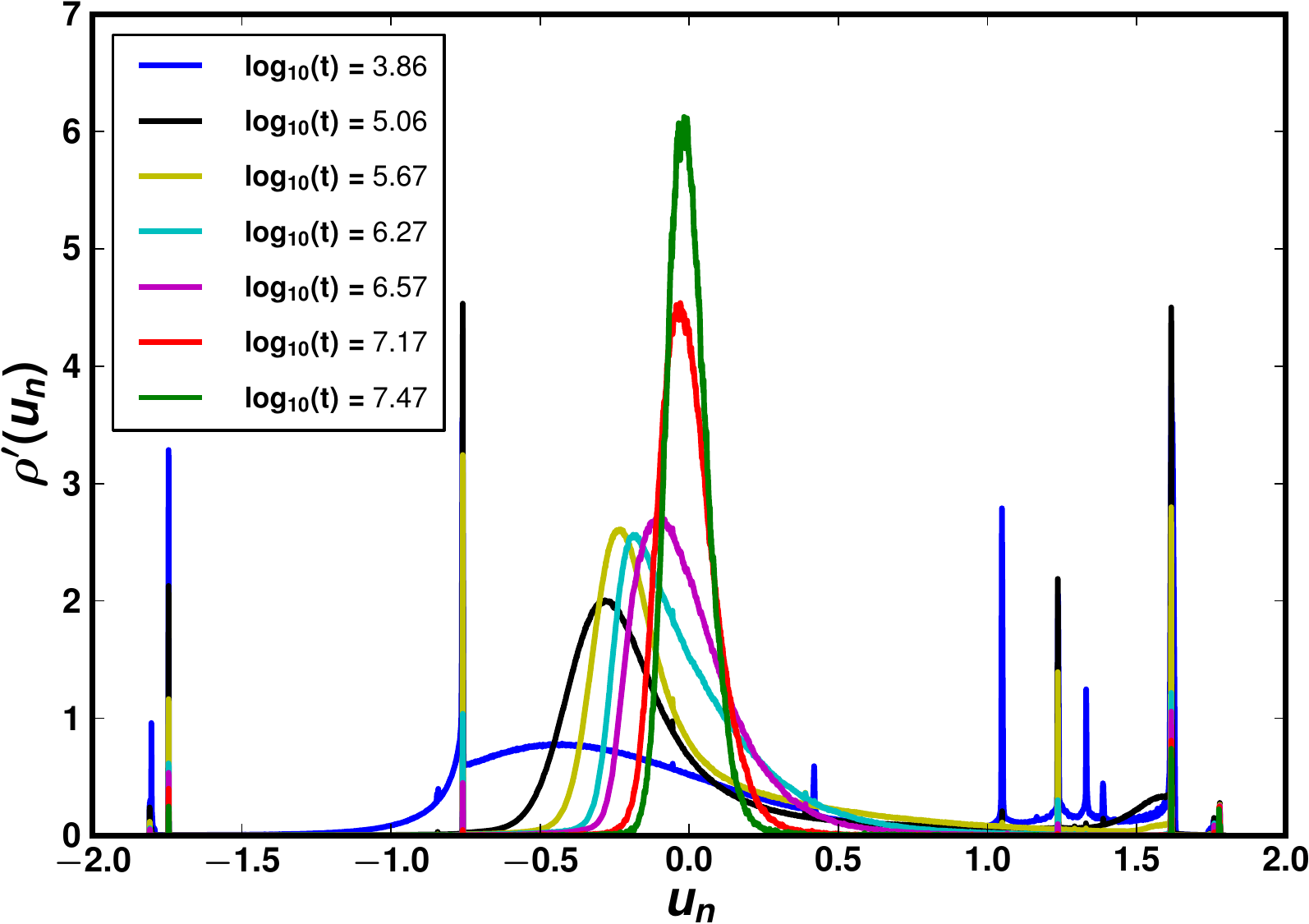}
\caption{ Distribution of average speed for the stochastic web with 
$\dot{\cX}'_0 = 1.39$ and $(\varepsilon, \beta^2)= (0.69, 1.83 \cdot 10^{-5})$ at 
$\omega_{\rmpe} \, t = 7 \cdot 10^3, 1.1 \cdot 10^5, 4.6 \cdot 10^5, 1.8 \cdot 10^6, 
3.7 \cdot 10^6$, $1.4 \cdot 10^7$ and $2.9 \cdot 10^7$.}
\label{arc_distrib_v0x_1p39}
\end{figure}
\begin{figure}[t]
\includegraphics[width=\linewidth]{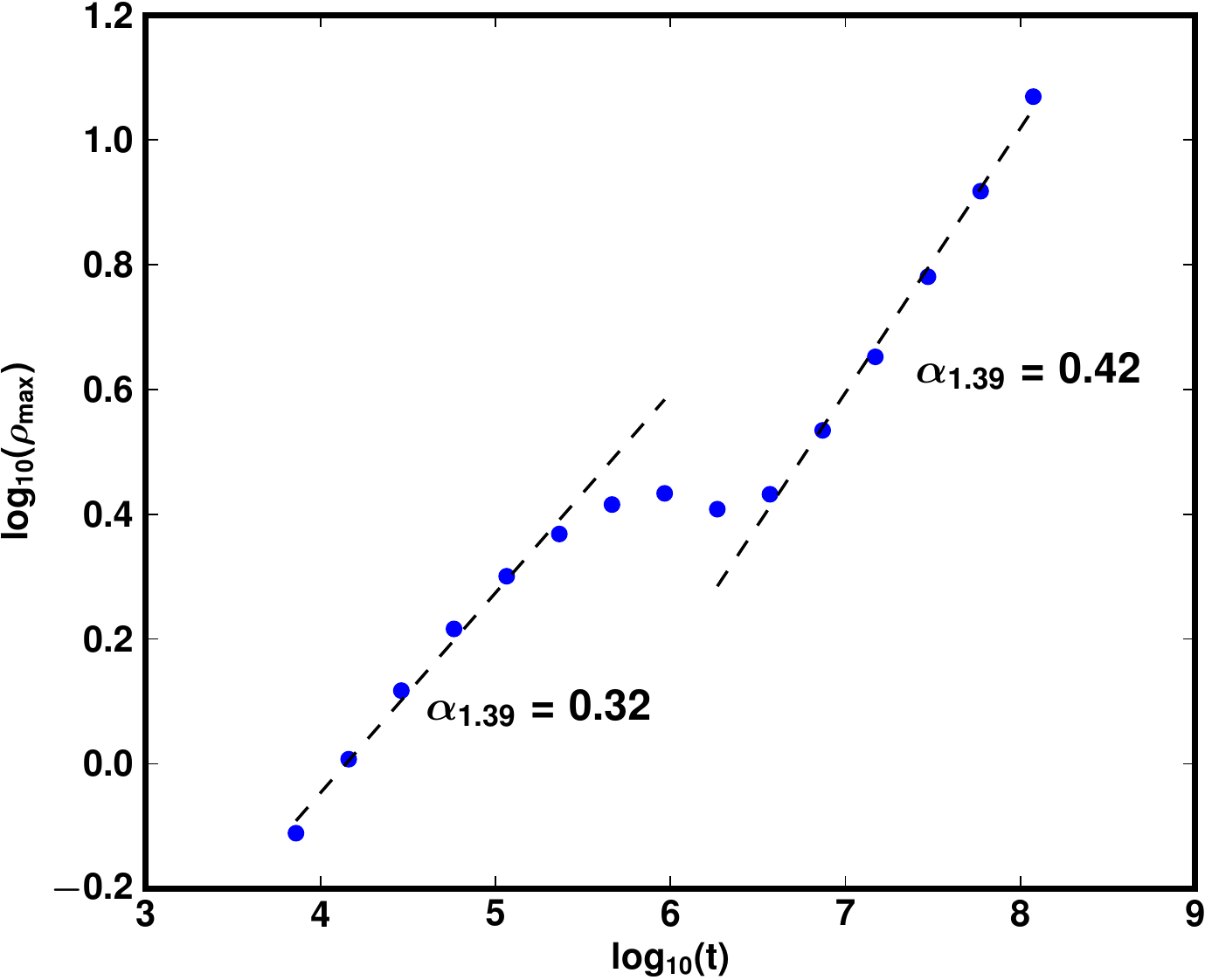}
\caption{Evolution of $\rho_{\max}$ versus $n$ for the time-independent 
Hamiltonian with $(\varepsilon, \beta^2)= (0.69, 1.83 \cdot 10^{-5})$ and $\dot{\cX}'_0 = 1.39$.}
\label{arc_max_v0x_1p39}
\end{figure}
%

%%%For Fig. 14-15, I choose $\dot{\cX'} = 1.39, \cX' = $random number between $[-\pi,\pi]$; 
%%%$\dot{cY'}$ = Gaussian weighted random number with standard deviation 1, 
%%%$\cY' =$ is chosen such that all the trajectory should be chaotic enough. 
%%%For that I use two random number-sets, one is between $[-0.1\pi : -0.6\pi];$ and $[+0.1\pi : +0.6\pi ]$. 
%%%We remove the points between $[-0.1\pi: 0.1\pi]$ 
%%%because in the Halloween mask-like phase space plot at the centre, there is a large region with regular trajectories. 
%%%For generating the distribution function one needs to check carefully, that all the trajectory should be chaotic enough. 
%%%Otherwise those regular trajectories create extra peaks. Therefore, one should choose the $\cY'$ value carefully. 
%%%And if some initial $\cY'$ values generate a regular trajectory 
%%%then I remove that number from the dataset and replace it with some other $\cY'$ value.     
{\it Transport in Halloween mask like web}:
For $\dot{\cX'}_0 = 1.39$ (further away from a simple rational) 
and $(\varepsilon, \beta^2)= (0.69, 1.83 \cdot 10^{-5})$,
we also draw 1024 initial conditions in the chaotic part of the domain 
defined by $-\pi \leq \cX'_0 \leq \pi$, 
$\dot{\cY'}_0$ a Gaussian random number with expectation 0 and standard deviation 1, 
and $\cY'_0$ outside the islands (typically, $0.1 \, \pi \leq | \cY'_0 | \leq 0.6 \, \pi$).
With these parameters and  initial conditions, 
the same analysis applies, as seen 
from the peaks in the distribution in Fig.~\ref{arc_distrib_v0x_1p39} 
and from the slopes $\alpha_{1.39} = (0.32, 0.42)$ in Fig.~\ref{arc_max_v0x_1p39}. 
In the next section, we discuss this change of exponent values more quantitatively.

%%%%%%%%%%%%%%%%%%%%%%%%%%%%%%%%%%%%%%%%%%%%%%%%%%%%%%%%%%%
\section{Effect of sticky island on anomalous transport: change of $\alpha$}
\label{sec:sticky_island}
%%%%%%%%%%%%%%%%%%%%%%%%%%%%%%%%%%%%%%%%%%%%%%%%%%%%%%%%%%%

%*** we do not mention the law of large numbers before this section ; 
%the reader may be surprised by the next sentence. 
%What do we want to point out ? 

In this section, we will attempt to quantify the impact of a sticky set on the transport, 
%A contrast between the law of large numbers and 
recalling that $\alpha$ must be 0.5 when the central limit theorem can be applied, 
whereas it is anomalous if $\alpha \ne 0.5$. 
The superdiffusive or subdiffusive nature of transport is measured by 
the weight of the tails in the distributions~: 
(i)~if there are fat tails (compared to the diffusive bulk of the distribution), 
the maximum $\rho_{\rm max}$ of the distribution will grow slower than power $1/2$ 
and we expect superdiffusion, 
while (ii)~if there are thinner tails, the maximum $\rho_{\rm max}$ of the distribution 
will grow faster than power $1/2$ and we expect subdiffusion. 
The  presence of the sharp peaks in the distribution of $\rho_n(\bar{v})$, 
Figs~\ref{arc_distrib}, \ref{arc_distrib_nu6}, \ref{arc_distrib_v0x_3}, 
\ref{arc_distrib_v0x_3p5}, \ref{arc_distrib_v0x_6} and \ref{arc_distrib_v0x_1p39}, 
increases the effective weight of the tail of the distribution, which makes exponent $\alpha < 1/2$. 

\begin{figure}
\includegraphics[width=\linewidth]{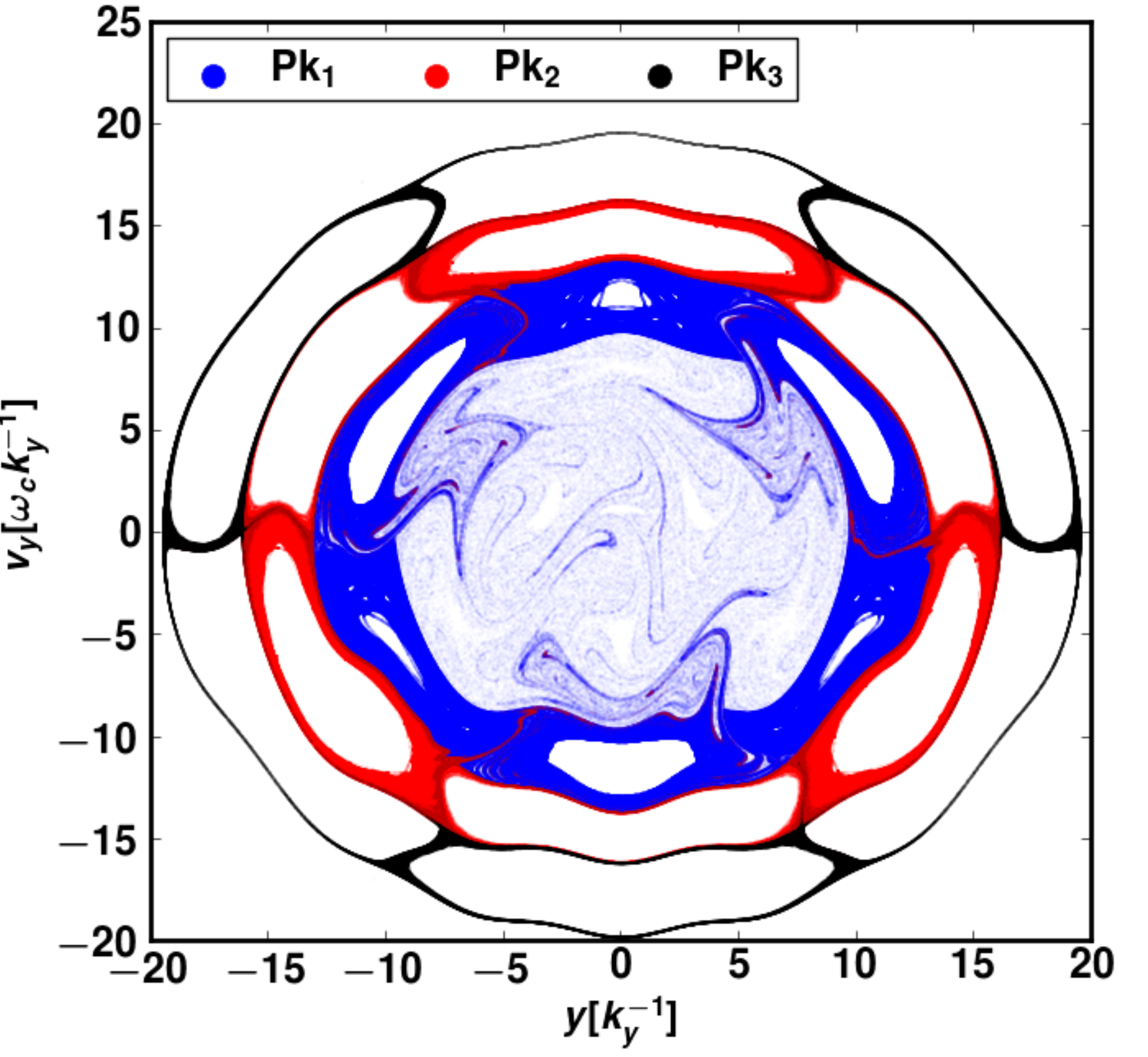}
\caption{Localization of three different sticky regions in the stochastic web
with three-fold symmetry of Eq.~(\ref{Auton_Red}) for $\varepsilon = 3.21$, 
$\beta^2 = 1.75 \cdot 10^{-6}$ and  ${\dot \cX_0}' = 3.0$ as in Fig.~\ref{web_vx_3_kx_0p001}. 
These different sticky regions are associated with three different peaks in the distribution
plot of average speed (Fig.~\ref{arc_distrib_v0x_3}) when $\log_{10}(\omega_{\rmpe} t)=2.92$,
which are
${\rm Pk_1}$ at $u_n = 1.0$ (blue dots), ${\rm Pk_2}$ at $u_n = 3.0$
(red dots) and ${\rm Pk_3}$ at $u_n = 5.0$ (black dots),
 respectively.}
\label{peaks_web_3fold_t0}
\end{figure}
Each of the sharp peaks is related to the presence of sticky islands in the 
phase space. Thus, one can select the portions of trajectories 
contributing to each peak and locate them in the Poincar\'e map. 
Specifically, given a trajectory $z_i(t)$ (with initial condition labeled 
$1 \leq i \leq N = 1024$) over a long time span $T_{\rm max}$,
and a sticking time $T_{\rm s}$, we fix a delay $\tau$, 
and consider the arc lengths 
$\Delta \cS'_{i, m}(T_{\rm s}) = \cS'_i(T_{\rm s} + m \tau) - \cS'_i(m \tau)$ 
of the portion of $z_i$ over $[m \tau, T_{\rm s} + m \tau]$, for 
$0 \leq m \leq M-1$ for some large $M= (T_{\rm max} -T_{\rm s})/\tau$,
where $T_{\rm max}$ is the maximum value of the time evolution,
$T_{\rm max} = 10^8$. 
For a moderate value of $\tau$ (say, $ \sim 200 $), 
%For a moderate value of $\tau$ (say, $ \sim 200 \, \omega_{\rmpe}^{-1}$), 
this method generates $M N$ time sequences of length $T_{\rm s}$ which 
we analyse. 
Indeed we generate the distribution functions using the same technique. 
Therefore, $T_{\rm s}$ is any time when the distribution function 
is generated. The minimum value of $T_{\rm s}$ and $\tau$ should be such that 
the dynamics become sufficiently ergodic within that time, 
i.e.\ it covers the entire chaotic domain of the phase space. 
Since we plot $T_s$ in $\log_{10}$ scale, 
we choose the values $T_s = 200, 400, 800, 1600, 3200, 6400 ...$. The 
maximum value of $T_s$ is determined by the relation $M \ge 50$, such that 
the total number of data points for each distribution $M N > 5\cdot 10^3$, 
which constructs a smooth distribution function at any time $T_s$.

In particular, we extract the points 
($\mathcal{X}', \mathcal{Y}', \mathcal{{\dot Y}}'$)
associated with each $\bar{v}_i$ that generates a sharp peak in the 
distribution and we plot their Poincar\'e section in the  
$(\cY', \rmd \cY' / \rmd t')$ plane, which generates the particular sticky set.
%In particular, when these portions keep sticking to the same island, 
%we plot their Poincar\'e section in the $(\cY', \rmd \cY' / \rmd t')$ plane for
%integer $\cX'/(2 \pi)$. 
This is done in Fig.~\ref{peaks_web_3fold_t0} for the web structures with 
three-fold 
symmetry for the trajectories with sticking duration $T_{\rm s}$ such that   
$\log_{10}(\omega_{\rmpe} T_{\rm s}) = 2.92$. 
There are three peaks, ${\rm Pk_1}$ at $u_n = 1.0$, ${\rm Pk_2}$ at 
$u_n = 3.0$ and ${\rm Pk_3}$ at $u_n = 5.0$, in Fig.~\ref{arc_distrib_v0x_3}, 
each with a finite width. 
For each peak, we identify the trajectories contributing to the peak, 
and plot their Poincar\'e section, whereby the sticky regions emerge. 
In Fig.~\ref{peaks_web_3fold_t0}, the sticky regions denoted by 
blue, red and black dots are associated with the peaks 
${\rm Pk_1}$, ${\rm Pk_2}$ and ${\rm Pk_3}$, respectively. 

\begin{figure}
\includegraphics[width=\linewidth]{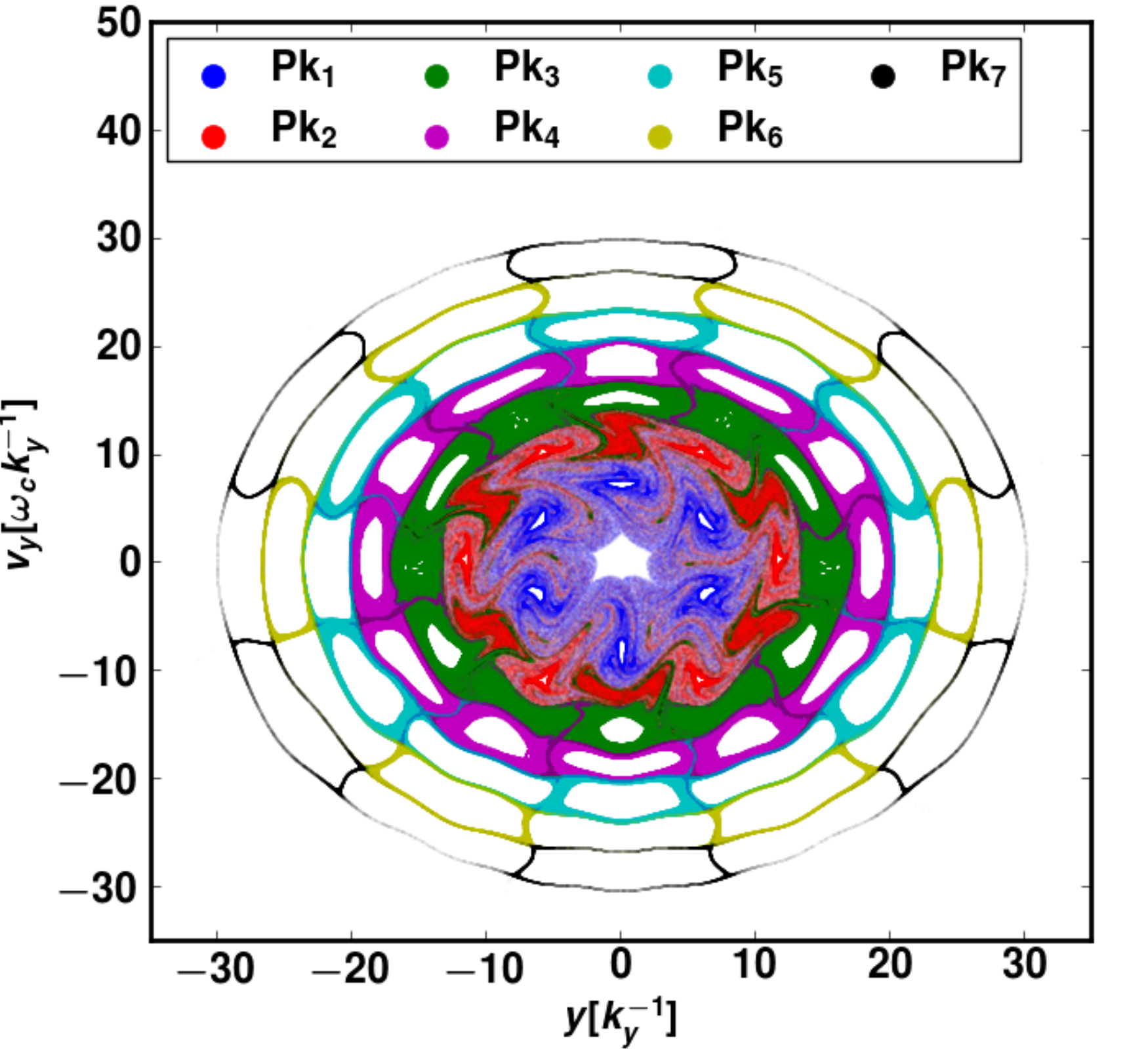}
\caption{Localization of seven different sticky regions in the stochastic web 
with six-fold symmetry of Eq.~(\ref{Auton_Red}) for $\varepsilon = 3.21$, 
$\beta^2 = 1.75 \cdot 10^{-6}$ and  ${\dot \cX_0}' = 6.0$. These different 
sticky regions are associated with seven different peaks in the distribution 
plot of average speed (Fig.~\ref{arc_distrib_v0x_6}) with sticking duration $T_{\rm s}$
such that $\log_{10}(\omega_{\rmpe} T_{\rm s}) = 2.92$, 
which are 
${\rm Pk_1}$ at $u_n = -3.6$ (blue dots), ${\rm Pk_2}$ at $u_n = -1.5$ 
(red dots), ${\rm Pk_3}$ at $u_n = 0.5$ (green dots), ${\rm Pk_4}$ at 
$u_n = 2.3$ (magenta dots), ${\rm Pk_5}$ at $u_n = 4.2$ (cyan dots), 
${\rm Pk_6}$ at $u_n = 6.2$ (yellow dots) and ${\rm Pk_7} = 8.1$ (black dots),
 respectively.}
\label{Peaks_web_6fold_t0}
\end{figure}

\begin{figure}
\includegraphics[width=\linewidth]{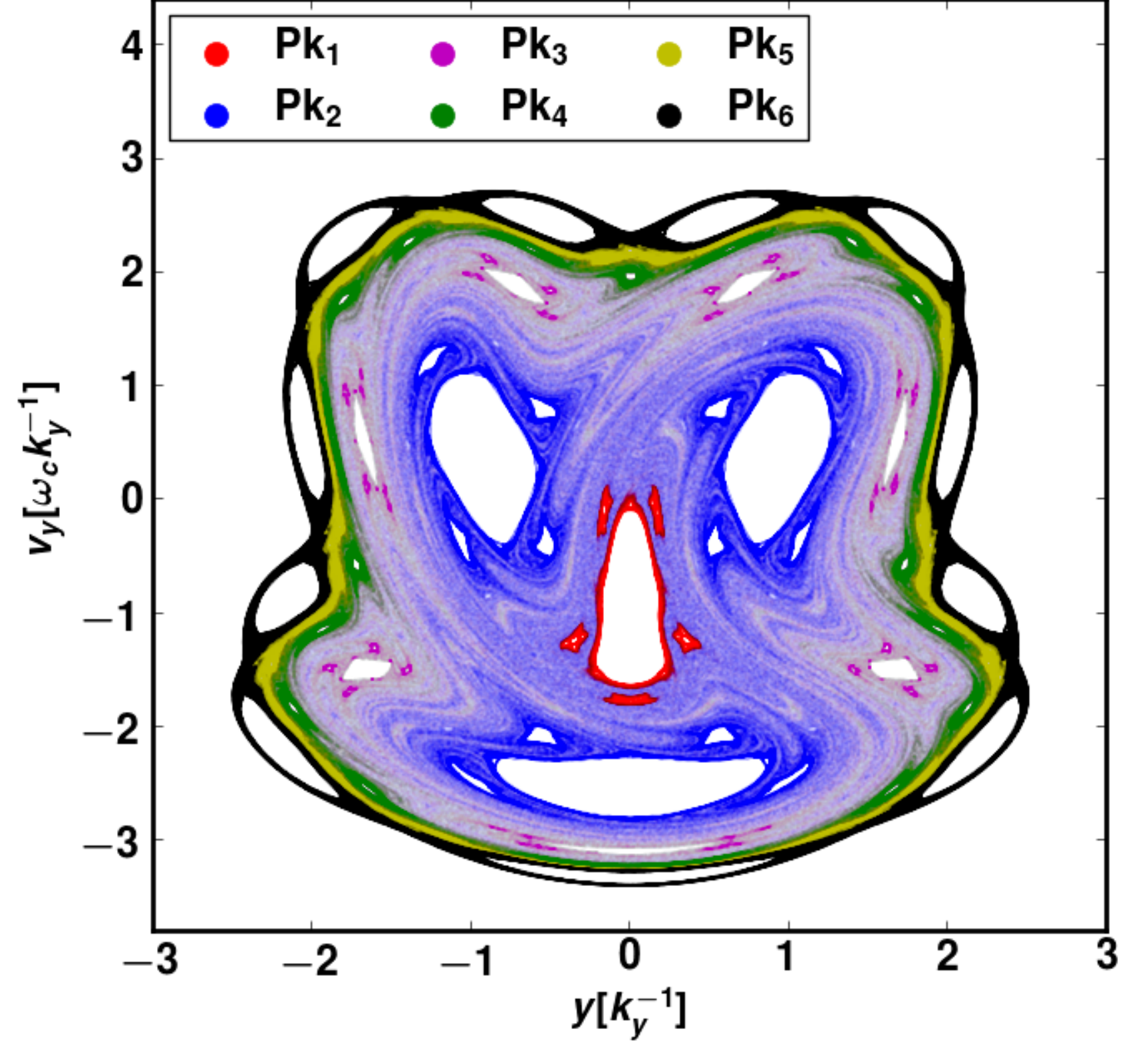}
\caption{Localization of six different sticky regions in the stochastic web 
with Halloween mask like structure of Eq.~(\ref{Auton_Red}) for 
$\varepsilon = 0.69$, 
$\beta^2 = 1.83 \cdot 10^{-5}$ and  ${\dot \cX_0}' = 1.39$. These different 
sticky regions are associated with six different peaks in the distribution 
plot of average speed (Fig.~\ref{arc_distrib_v0x_1p39}) with sticking duration $T_{\rm s}$
such that $\log_{10}(\omega_{\rmpe} T_{\rm s}) = 3.86$, 
which are 
${\rm Pk_1}$ at $u_n = -1.75$ (red dots), ${\rm Pk_2}$ at $u_n = -7.5$ 
(blue dots), ${\rm Pk_3}$ at $u_n = 0.4$ (magenta dots), ${\rm Pk_4}$ at 
$u_n = 1.05$ (green dots), ${\rm Pk_5}$ at $u_n = 1.22, 1.33, 1.38$ (yellow dots), 
${\rm Pk_6}$ at $u_n = 1.6, 1.78$ (black dots),
 respectively.}
\label{Peaks_web_1p39fold_t0}
\end{figure}

In Fig.~\ref{arc_distrib_v0x_6}, the distribution of average speed for the web 
with six-fold symmetry has seven peaks, namely ${\rm Pk_1}$ at $u_n = -3.6$, 
${\rm Pk_2}$ at $u_n = -1.5$, ${\rm Pk_3}$ at $u_n = 0.5$, ${\rm Pk_4}$ at
$u_n = 2.3$, ${\rm Pk_5}$ at $u_n = 4.2$, ${\rm Pk_6}$ at $u_n = 6.2$
 and ${\rm Pk_7}$ at $u_n = 8.1$. In a similar way, we locate the sticky 
region in the phase space for each of these peaks for the same duration $T_{\rm s}$,
such that $\log_{10}(\omega_{\rmpe} T_{\rm s}) = 2.92$. 
In Fig.~\ref{Peaks_web_6fold_t0}, the blue, red, green, magenta, cyan, yellow and black dots 
identify the sticky regions associated with the peaks ${\rm Pk_1}$ to ${\rm Pk_7}$, respectively. 
Thus, all peaks in the distribution plots are associated with different sticky sets. In Fig.~\ref{Peaks_web_1p39fold_t0}, 
we similarly identify the sticky sets associated with the peaks for the stochastic web 
with Halloween mask like structure of Eq.~(\ref{Auton_Red}) for 
$\varepsilon = 0.69$, 
$\beta^2 = 1.83 \cdot 10^{-5}$ and  ${\dot \cX_0}' = 1.39$.

Sticky sets are special in the dynamics, because they influence transport in the chaotic domain. 
Therefore, they are not islands in which the trajectory remains forever : 
their trajectories leave them to wander further through the chaotic domain. 
This implies that a group of trajectories initially in the sticky set will leak into the main chaotic sea 
-- but slowly enough for their stickiness to show up. 
Geometrically, sticky regions correspond to tangles bordering islands.
If the initial condition of a particle is well inside an island, the dynamics of 
the particle will be essentially regular~: it never exits from the islands chain and 
generates a sharp peak in the distribution which never becomes empty with time.
Those regular trajectories do no contribute to the anomalous 
transport. We exclude all the regular trajectories to remove those 
peaks by considering all the initial condition inside the chaotic domain. 

Each sticky set \emph{leaks} into the main chaotic domain 
at its own rate, which we discuss below. 
However, because our dynamics is hamiltonian, 
we also know from the Poincar\'e recurrence theorem that any trajectory will,
after a sufficiently long but finite time, return to a state arbitrarily
close to its initial state. Therefore, the leakage from a sticky set is a transient process 
and, in general, the trajectories having left such a set behave afterwards 
like any other trajectory of the chaotic domain.  
It will almost surely return, at a later time, close to its initial state, 
but the statistics of this Poincar\'e recurrence time and its sensitivity to the initial data 
are another issue, which we do not discuss. 

%***
%If the initial condition is chosen inside the chaotic domain of the phase space, 
%during the time evolution it may stick near a regular trajectory 
%and spend a longer time there than in the other region, 
%but finally it will come back close to its initial conditions.
%The sticky regions correspond to the tangle bordering the island.
%If the initial condition of a particle is well inside an island, the dynamics of 
%the particle will be essentially regular~: it never exits from the islands chain and 
%generates a sharp peak in the distribution which never becomes empty with time.
%Those regular trajectories do no contribute to the anomalous 
%transport. We exclude all the regular trajectories to remove those 
%peaks by considering all the initial condition inside the chaotic domain. 
%Therefore, due to the Poincar\'e recurrence theorem, 
%each of the sticky sets becomes empty as time goes on and contribute 
%to increase the
%maximum of the average speed distribution, but some sticky sets may {\it leak}
%slower or faster than others, which can change the transport coefficient.
%*** 

To understand the influence of stickiness on anomalous transport,
we now consider the web structure with three-fold symmetry, 
and investigate the change of each peak for increasing evolution time 
$t = n \Delta t$. 
This analysis makes it possible to see the time evolution of the particles 
trapped in the corresponding islands. 
From the distribution $\rho_n(\bar{v})$ of average speed, 
one can count the number of data points that contribute to
each specific peak at different times $t$. Then one estimates 
how long the particles are sticking to each specific island, by monitoring the
change of area localized under each of those peaks as a function of $n$. 
Therefore, this area yields the weight of sticking to that particular
island, until at least $t_1 = n \Delta t$, which can be written as
\begin{eqnarray}
  w_{\rm Pk}(t_1, T_{\max{}}) 
  = (T_{\max{}}-t_1)^{-1} \int_{t_1}^{T_{\max{}}}\rho_{\rm Pk}(t) \rmd t ,
\label{Weight_stick}
\end{eqnarray}
where ${\rm Pk}$ is the index for each peak, $\rho_{\rm PK}(t)$ is the area
under the peak ${\rm Pk}$ at time $t$, and the statistics are gathered from a 
``very long'' run $[0, T_{\max{}}]$. 
Under an ergodic assumption \cite{leoncini04, Meziani:b2012}, 
this weight would enable one to estimate the probability 
that a trajectory would stick to island ${\rm Pk}$ for at least the duration $t_1$. 
%in the form  $\lim_{T_{\max{}} \to \infty}   w_{\rm Pk}(\tau, T_{\max{}}) / w_{\rm Pk}(0, T_{\max{}})$. 
For large sticking time, 
a self-similar behaviour in the small scales in phase space near the island 
will be associated with a power law decay with an exponent $\gamma$, 
\begin{eqnarray}
  w_{\rm Pk}(T_{\rm s}, T_{\max}) 
  \sim T_{\rm s}^{1 - \gamma_{\rm Pk}} ,
\label{stick_expo}
\end{eqnarray}
where we consider $\rho_{\rm Pk}(t) \propto t^{-\gamma}$ and replace $t_1$ with 
$T_{\rm s}$. This integration is valid only for $\gamma_{\rm PK} > 1$.
%In order to analyze the sticking-times statistics, we count the number of data
%points sticking to each of the islands and plot them in 
%logarithmic scale versus the duration. 
%For counting data points, we consider three different 
%annular areas in phase space, one for each peak, 
%using the radius $\cR' = \sqrt{{\cY'}^2 + (\rmd \cY' / \rmd t')^2}$  
%related to the gyration adiabatic invariant. 
%%which would be a constant of the motion for the cyclotron dynamics of $\cY'$ if $\varepsilon$ were zero. 
%%Actually, $r^2$ is (up to the normalizing factor $\omega_\rmc$ set to 1 in reduced time units) the action, 
%%which generates an adiabatic invariant for small $\varepsilon$. 

In order to analyze the sticking-times statistics, 
we count the number of data points sticking to each island 
and plot them in logarithmic scale versus the duration. 
One way to do is by integrating the distribution along the $u$ axis, 
at the particular sharp peak location. 
But when we locate, in the ($\mathcal{Y}', \mathcal{{\dot Y}}'$) Poincar\'e section, 
the dots associated with the sharp peaks, 
we found some dots actually situated within the main chaotic domain, and 
moreover the sticky sets overlap with each other on this one-variable ($u$) axis. 
Therefore, to find the actual $\rho_{\rm{Pk}}$ value associated with the sticky sets only, 
one should count the dots situated only near the chaotic tangle bordering the island.
Each sticky set leaks with time and mixes into the more regular chaotic domain.
%
%On the one hand, in the distribution function, their sharp peaks are separated 
%along the $u$ axis; on the other hand, in the $(\cY', \rmd \cY' / \rmd t')$ section plane, those
%sticky sets overlap with each other. 
%

So, we define the border of the web network 
by an invariant surface with dimensionless radius
$\cR'  = \sqrt{{\cY'}^2 + (\rmd \cY' / \rmd t')^2}$, which is
associated with the gyration action.
The number density within each sticky set 
decreases rapidly away from the surface $\cR'$ = constant.
%
%with the dimensionless radius 
%$\cR'  = \sqrt{{\cY'}^2 + (\rmd \cY' / \rmd t')^2}$  
%related to the gyration adiabatic invariant. 
%
Therefore, to count the dots belonging only to the sticky sets, 
we denote the maximum and minimum value of $\cR'$ 
by $\cR'_{\rm out}$ and $\cR'_{\rm in}$ respectively for each peak. 
We consider three different annular domains in phase space, one for each peak, 
using the radius $\cR'$. 
%which would be a constant of the motion for the cyclotron dynamics of $\cY'$ if $\varepsilon$ were zero. 
%Actually, $r^2$ is (up to a normalizing factor) the action, 
%which generates an adiabatic invariant for small $\varepsilon$.
In the $(\cY', \rmd \cY' / \rmd t')$ section plane, ${\rm Pk}_1$ is associated with the annulus 
with inner radius $\cR'_{\rm in} = 9.4$ and outer radius $\cR'_{\rm out} = 13.5$~; 
similarly, ${\rm Pk}_2$ with $\cR'_{\rm in} = 12.23$ and $\cR'_{\rm out} = 16.76$, and 
${\rm Pk}_3$ with $\cR'_{\rm in} = 15.72$ and $\cR'_{\rm out} = 20.0$. 
From the data set associated with each peak, 
we identify those points which satisfy 
$\cR'_{\rm in} \le k_y \sqrt{\cY^2 + {\dot \cY}^2 / \omega_\rmc^2} \le \cR'_{\rm out}$. 
We perform this counting for all the $n$ values to obtain $W_{\rm Pk}(n) \cong M N w_{\rm Pk}$, 
we plot $\log_{10}(W_{\rm Pk})$ vs. $\log_{10}(\omega_{\rmpe} T_{\rm s})$, 
and read the exponent $\gamma_{\rm Pk}$ from the slope according to Eq.~(\ref{stick_expo}). During the counting process, one should keep in mind that the 
length of analysed data set $M N$ should be constant for all the time $T_s$. 
%One way for doing this is by
%changing the length $T_{\rm max}$ or $\tau$ for different $T_s$ such that
%$M$ always remain constant.
%
\begin{figure}
\includegraphics[width=\linewidth]{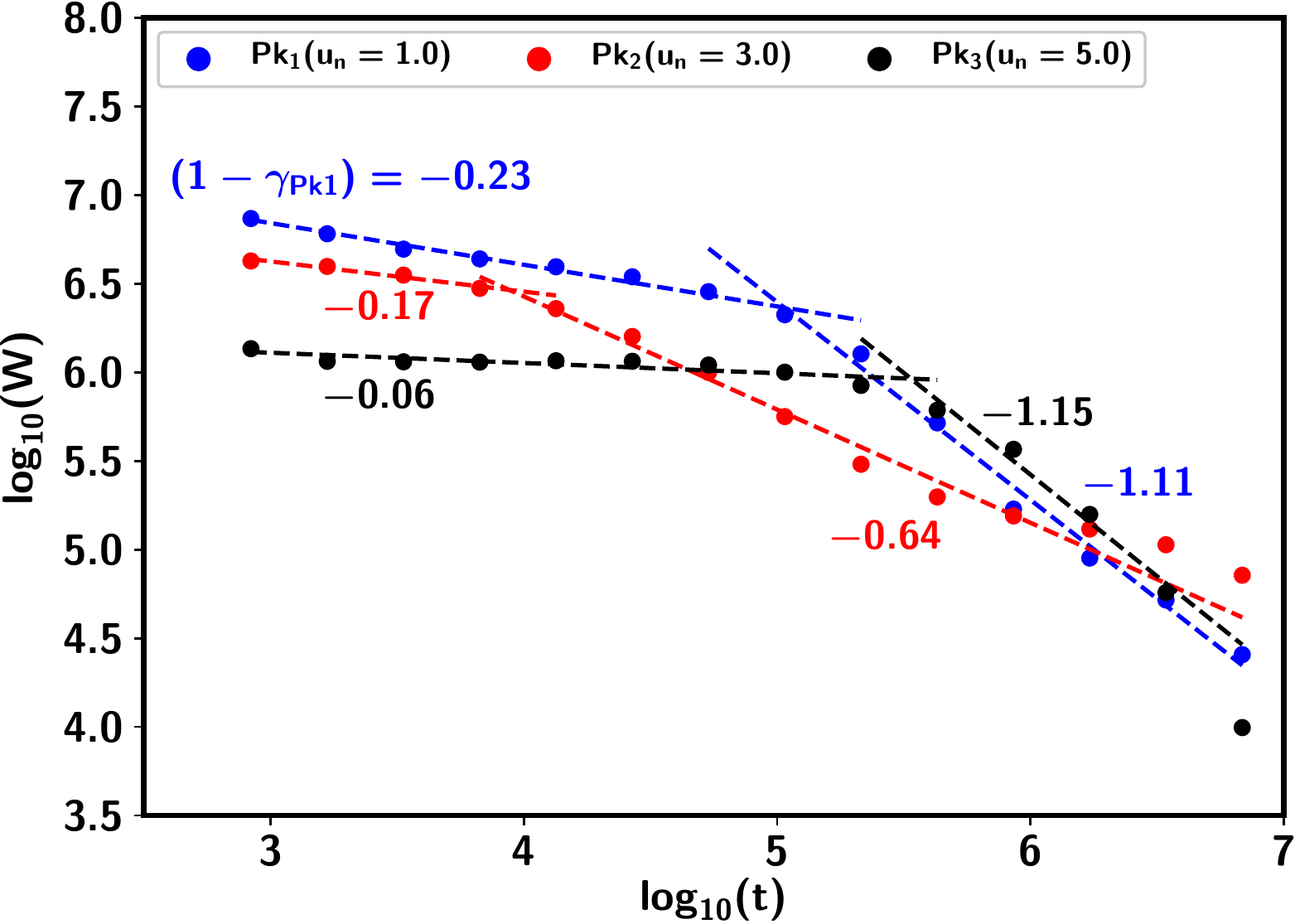}
\caption{Time evolution of number of points $W_{\rm Pk}$ in each of the 
three peaks 
${\rm Pk_1}$ at $u_n = 1.0$ (blue dots), ${\rm Pk_2}$ at $u_n = 3.0$
(red dots) and ${\rm Pk_3}$ at $u_n = 5.0$ (black dots), respectively,
in the distribution of average speed (Fig.~\ref{arc_distrib_v0x_3}). 
Here $t$ denotes the sticking time $T_{\rm s}$.}
\label{time_evo_peaks_v0x3}
\end{figure}

Fig.~\ref{time_evo_peaks_v0x3} presents these results. 
The time evolutions of ${\rm Pk_1}$, ${\rm Pk_2}$ and ${\rm Pk_3}$ are presented 
by blue, red and black dot respectively. Among all three peaks, 
${\rm Pk_1}$ is the strongest peak in the distribution. 
Initially, the exponents in Eq.~(\ref{stick_expo}) for all three peaks 
have very small values $|1 - \gamma_{\rm Pk}| \ll 1$, 
namely $\gamma_{\rm Pk1} = 1.23$,  $\gamma_{\rm Pk2} = 1.17$ and $\gamma_{\rm Pk3} = 1.06$. 
For $\log_{10}(\omega_{\rmpe} T_{\rm s}) = 5.3$, the exponent value for the strongest peak ${\rm Pk_1}$ 
increases to $\gamma_{\rm Pk1} = 2.11$.
The cross-over of the second strongest peak ${\rm Pk_2}$ occurs for $\log_{10}(\omega_{\rmpe} T_{\rm s}) = 4.2$,
when the exponent value changes to $\gamma_{\rm Pk2} = 1.64$, which still implies $|1 - \gamma_{\rm Pk2}| < 1$.
For the weakest peak ${\rm Pk_3}$, the exponent changes to 
$\gamma_{\rm Pk3} = 2.15$ when $\log_{10}(\omega_{\rmpe} T_{\rm s}) = 5.7$. 
Since the exponent value for the strongest peak changes when $\log_{10}(\omega_{\rmpe} T_{\rm s}) = 5.3$,
the strength of this peak starts to decrease (leak) faster, 
which helps to increase the maximum $\rho_{\rm max}$ of the average distribution 
$\rho_n(\bar{v})$ at a faster rate. 
Therefore, the value of the transport exponent $\alpha$, from Eq.~(\ref{arc_length_max}), 
increases after $\log_{10}(\omega_{\rmpe} T_{\rm s}) = 5.3$, which is also observed in Fig.~\ref{arc_max_tind}.

\begin{figure}
\includegraphics[width=\linewidth]{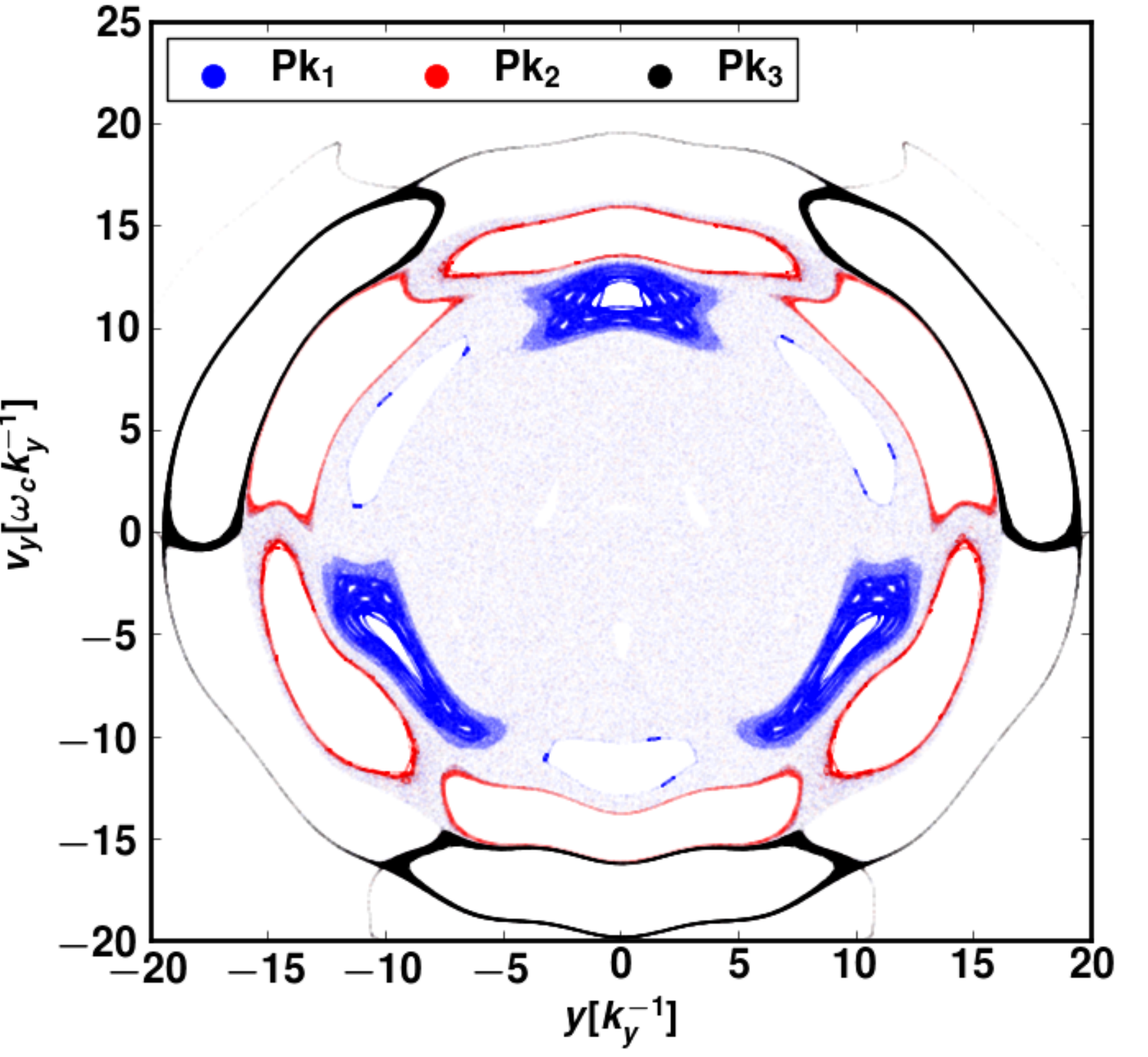}
\caption{ Three sticky regions in the stochastic web
with three-fold symmetry for $\varepsilon = 3.21$, 
$\beta^2 = 1.75 \cdot 10^{-6}$ and  ${\dot \cX_0}' = 3.0$ as in Fig.~\ref{peaks_web_3fold_t0} 
with sticking duration $T_{\rm s}$
such that $\log_{10}(\omega_{\rmpe} T_{\rm s}) = 5.63$.}
\label{peaks_web_3fold_t10}
\end{figure}

Fig.~\ref{peaks_web_3fold_t10} presents the three sticky regions for the sticking time $T_{\rm s}$ 
such that $\log_{10}(\omega_{\rmpe} T_{\rm s}) = 5.63$. 
Comparing with Fig.~\ref{peaks_web_3fold_t0}
(with $\log_{10}(\omega_{\rmpe} T_{\rm s}) = 2.92$)
extracted from the same $M N$ time sequences, 
we see that the strength (number of dots) 
of the sticky set associated with the peak ${\rm Pk_1}$ decreases 
by a very large amount and starts to become empty. 
%The number of dots in the three sticky sets change from 7377159 to 518394 for ${\rm Pk_1}$, 
%from 4249939 to 198201 ${\rm Pk_2}$ and from 1359991 to 611354 ${\rm Pk_3}$, respectively.  

%%%%%%%%%%%%%%%%%%%%%%%%%%%%%%%%%%%%%%%%
\section{Conclusions}
\label{sec:Conclusion}
%%%%%%%%%%%%%%%%%%%%%%%%%%%%%%%%%%%%%%%%

In this paper, we discuss the transport due to electrostatic waves generated by the 
$\bfE \times \bfB$ electron drift instability. 
The original time-dependent 3-degrees-of-freedom problem is reduced to 
a 2-degrees-of-freedom time-dependent model 
and a 2-degrees-of-freedom autonomous model. 
Due to the wave-particle interaction, the dynamics become chaotic, and
trajectories form stochastic web structures with different shape for different parameters,
which we investigated for both the time-dependent and time-independent descriptions. 
%For a particular set of parameter we generate a Halloween mask  
%like stochastic web structures. 

Along with each web structure, there occur sticky islands 
where the trajectory spends more time compared to the purely chaotic domain, 
which affects the diffusion rate \cite{mandal19a}. 
We use a scaling exponent for characterising the particle transport, 
and find that the transport is anomalous, of super-diffusive type.
The presence of sticky islands generates sharp peaks in the 
distribution of average speed (a phase space observable) which 
increases the effective weight of the tail in the distribution. 
Considering the Poincar\'{e} recurrence theorem and Kac' lemma for the sticky sets, 
we estimate a power law decay for the probability that a trajectory would stick to an  
island. 
With increasing duration of the time evolution, 
sticky sets start to become empty and they decay with a higher exponent value. 
This change in the exponent $\gamma$ values also affects the 
transport-coefficient exponent $\alpha$ values.

In real Hall thrusters, the $\bfE \times \bfB$ instability generates 
many unstable modes, with different frequencies and wavevectors. 
In this case, even for small amplitude waves, 
the dynamics cannot be reduced to a time-independent 2-degrees-of-freedom model. 
However, each wave will typically bear its own dimensionless parameters 
$(\varepsilon_i, \beta_i, k_{zi}/k_{yi})$,
with small values for $\beta_i$ and $k_{zi}/k_{yi}$. 
Therefore, the several-wave dynamics will exhibit 
resonance overlap between the structures generated by these individual waves, 
resulting in smaller islands (if any survive \cite{ElsEsc91, ElsEsc93, Neishtadt19}) 
and more regular transport \cite{mandal19a}. 
In case of presence of harmonics modes, 
different types of web structures (mixed phase space) are formed in the phase space
\cite{vasilev91}, which therefore will generate super-diffusive transport. 

Beside the effect of several waves, three issues must also be considered. 
First, the thruster chamber is a cylinder, 
where the intensity of the radial magnetic field decreases for larger radius (here $x$) 
and the azimuthal coordinate (here $y$) is periodic. 
Second, electrons do not stay forever in the chamber, 
which implies that tools of transient chaos \cite{Tel15} will be relevant. 
Third, the electrons charge and current generate electromagnetic fields,
so that the system needs a self-consistent many-body description.

%%%%%%%%%%%%%%%%%%%%%%%%%%%%%%%%%%%%%%%%
\section*{Acknowledgements}
%%%%%%%%%%%%%%%%%%%%%%%%%%%%%%%%%%%%%%%%
This work is part of IFCPRA project 5204-3. 
We acknowledge the financial support from CEFIPRA/IFCPRA. 
The Centre de Calcul Intensif d'Aix-Marseille is acknowledged 
for granting access to its high performance computing resources. 
We are grateful to Professors Dominique Escande and Abhijit Sen 
for many fruitful discussions, 
and to anonymous reviewers for their comments.

%%%%%%%%%%%%%%%%%%%%%%%%%%%%%%%%%%%%%%%%%%%
%\begin{wrapfigure}{r}{70mm}\centering
%\vspace{0cm} % Adjust vertical figure placement
%\includegraphics[width=70mm]{epslogo}
%\caption{\it \small EPS logo}
%\label{fig:flow}
%\vspace{0cm} % Adjust vertical figure spacing
%\end{wrapfigure}

\end{document}